\documentclass[12pt,oneside, a4paper]{article}
\pdfoutput=1

\ifx\pdfoutput\undefined
\usepackage[dvips,bookmarks=false]{hyperref}	
\else
\usepackage{hyperref}	
\fi
\hypersetup{colorlinks,bookmarksopen,bookmarksnumbered,citecolor=blue,
linkcolor=black,pdfstartview=FitH,urlcolor=blue}

\usepackage{graphicx}
\usepackage{amssymb}
\usepackage{cite}
\usepackage{bm}
\usepackage{indentfirst}
\usepackage{amsmath}
\usepackage{hhline}

\newcommand{\capdef}{}
\newcommand{\mycaption}[2][\capdef]{\renewcommand{\capdef}{#2}%
       \caption[#1]{{\footnotesize #2}}}

\begin{document}

\begin{titlepage}

\begin{flushright}
KANAZAWA-11-14
\end{flushright}

\begin{center}

\vspace{1cm}
{\large\bf Direct Detection of Leptophilic Dark Matter  
in a Model with\\[2mm]
 Radiative Neutrino Masses}
\vspace{1cm}

\renewcommand{\thefootnote}{\fnsymbol{footnote}}
Daniel Schmidt$^{1,}$\footnote[1]{daniel.schmidt@mpi-hd.mpg.de}, 
Thomas Schwetz$^{1,}$\footnote[2]{schwetz@mpi-hd.mpg.de}
and
 Takashi Toma$^{1, 2,}$\footnote[3]{t-toma@hep.s.kanazawa-u.ac.jp}
\vspace{5mm}

{\it%
$^{1}${Max-Planck-Institut f\"ur Kernphysik, Saupfercheckweg 1, 69117 Heidelberg, Germany}\\
$^{2}${Institute for Theoretical Physics, Kanazawa University, Kanazawa
  920-1192, Japan}}

\vspace{8mm}

\abstract{We consider an electro-weak scale model for Dark Matter (DM) and
  radiative neutrino mass generation. Despite the leptophilic nature
  of DM with no direct couplings to quarks and gluons, scattering with
  nuclei is induced at the 1-loop level through photon exchange.
  Effectively, there are charge-charge, dipole-charge and
  dipole-dipole interactions. We investigate the parameter space
  consistent with constraints from neutrino masses and mixing, charged
  lepton-flavour violation, perturbativity, and the thermal production
  of the correct DM abundance, and calculate the expected event rate
  in DM direct detection experiments. We show that current data from
  XENON100 start to constrain certain regions of the allowed
  parameter space, whereas future data from XENON1T has the potential
  to significantly probe the model.
}

\end{center}
\end{titlepage}

\renewcommand{\thefootnote}{\arabic{footnote}}
\setcounter{footnote}{0}

\setcounter{page}{2}

\section{Introduction}

The Standard Model (SM) is very successful in describing the
fundamental particles of our world. The only solid evidence for its
failure so far is the fact that neutrinos have mass, which is a
necessity due to the observation of neutrino
oscillations~\cite{Fukuda:1998mi, Ahmad:2002jz, Araki:2004mb,
  Adamson:2008zt}. Furthermore, the standard cosmological model, the
$\Lambda$CDM model, provides an excellent description of our Universe,
with the exception that within the SM there is no viable candidate for
a Dark Matter (DM) particle, which is an important ingredient of the
$\Lambda$CDM model, supported by observations such as the rotation
curves of spiral galaxies~\cite{Begeman:1991iy}, WMAP CMB
measurements~\cite{Komatsu:2010fb} and gravitational
lensing~\cite{Massey:2007wb}. Hence, both neutrinos, as well as DM
require an extension of the SM. Often these two phenomena are
considered separately, since they might be manifestations of physics
from vastly different energy scales. Here we adopt the hypothesis that
neutrino mass and DM are related, and both emerge from physics at the
TeV scale. In this respect models which generate neutrino masses
radiatively~\cite{Zee:1980ai, Zee:1985rj, Ma:2006km, Zee:1985id,
  Babu:1988ki, Krauss:2002px, Aoki:2008av} are intriguing.\footnote{This
  is only a small collection of references. There are many more models
  which generate neutrino masses radiatively, also including
  supersymmetry, see e.g.~\cite{ArkaniHamed:2000kj, Hirsch:2000ef, Ma:2006uv}.}
Loop suppression factors and several powers of Yukawa couplings can bring
the scale of neutrino mass generation down to the TeV, and symmetries
required to stabilize DM may play a role for neutrinos, for example
forbid tree-level mass terms. Recent works in this context can be
found in refs.~\cite{Sahu:2008aw, Kubo:2006yx, Kajiyama:2006ww,
  Suematsu:2009ww, Suematsu:2010gv, Sierra:2008wj, Babu:2007sm,
  Suematsu:2010nd, Fukuoka:2010kx, Aoki:2010tf, Aoki:2010ib,
  Lindner:2011it, Li:2010rb, Kanemura:2011jj, Ibarra:2011gn,
  Kajiyama:2011fe, Kajiyama:2011fx, Aoki:2011he, Kanemura:2011mw,
  Kanemura:2011vm, Aoki:2011yk, Farzan:2010mr}.

The so-called WIMP hypothesis suggests that DM interacts sufficiently with
SM particles in order to generate the relevant abundance due to thermal
freeze-out from the primordial plasma. This motivates the direct search for
DM in our galaxy by looking for the scattering of DM particles with nuclei
in underground detectors. Several direct detection experiments are pursuing
such searches, for example the CDMS II~\cite{Ahmed:2009zw},
XENON100~\cite{Aprile:2010um, Aprile:2011hi}, CoGeNT~\cite{Aalseth:2010vx},
DAMA/LIBRA~\cite{Bernabei:2010mq}, CRESST-II~\cite{Angloher:2008jj},
ZEPLIN-III~\cite{Lebedenko:2008gb} and KIMS~\cite{Lee.:2007qn, kims-taup}
experiments. In typical WIMP models DM interacts directly with quarks,
providing DM--nucleus scattering at tree-level~\cite{Jungman:1995df,
Belanger:2008sj}. Here we are interested in so-called ``leptophilic''
models, where DM couples directly only to leptons, see
e.g.~\cite{arXiv:0811.0399}. Even in that case, DM--nucleus interactions can
be induced at loop-level due to the exchange of the
photon~\cite{Kopp:2009et}. The resulting effective interactions have been
investigated in refs.~\cite{Kopp:2009et, Agrawal:2011ze}. In the following
we will consider a model where the corresponding loop-diagrams induce a
magnetic and/or electric dipole moment interaction~\cite{Pospelov:2000bq,
Sigurdson:2004zp, Masso:2009mu, Chang:2010en, Barger:2010gv, Cho:2010br,
Fitzpatrick:2010br, arXiv:1007.5515}.

We consider a model proposed by Ma~\cite{Ma:2006km}, in
which neutrino masses are generated through 1-loop interactions and
the particles which propagate in the loop can be DM candidates, being
leptophilic by construction. The DM phenomenology of the model and 
extended versions thereof has been studied in refs.~\cite{Kubo:2006yx,
  Kajiyama:2006ww, Suematsu:2009ww, Suematsu:2010gv, Sierra:2008wj,
  Kajiyama:2011fe, Kajiyama:2011fx} and prospects for collider
searches have been studied in refs.~\cite{Sierra:2008wj, Aoki:2010tf}.
We consider the situation that the lightest right handed neutrino is
the DM candidate and the second lightest right handed neutrino is
almost degenerated with the DM candidate. Under this situation,
elastic DM--nucleus scattering is extremely suppressed and inelastic
scattering induced by a lepton-loop coupled to the photon gives the
dominant contribution to the event rate in direct detection
experiments. We calculate the event rate in the model and compare it
with XENON100, KIMS and DAMA data. The paper is organized as
follows. In Section~2, we shortly review the model from
ref.~\cite{Ma:2006km}. We discuss the constraints from neutrino
oscillation data, lepton-flavour violation and the thermal production
of the DM relic abundance. In Section~3, we discuss the inelastic
scattering cross section in an effective theory approach and calculate
the event rate.  Moreover, monochromatic photons from the decay of the
excited DM state are also discussed.  We summarize and conclude in
Section~4. Explicit functions needed in the effective theory approach
are listed in the Appendix~A.


\section{The Model}

\subsection{Neutrino masses and mixing}

The model proposed by Ma in ref.~\cite{Ma:2006km} is a simple extension of
the SM, which correlates neutrino physics and the existence of DM. The added
particles to the SM are three right handed neutrinos $N_i$ $(i=1,2,3)$ and
one inert Higgs doublet $\eta$.  In addition, a discrete $\mathbb{Z}_2$
symmetry is imposed: odd for the new particles and even for SM particles.
The new invariant Lagrangian is 
\begin{equation}
\mathcal{L}_N=\overline{N_i}i\partial\!\!\!/\!\:P_RN_i
+\left(D_\mu\eta\right)^\dag\left(D^\mu\eta\right)
-\frac{M_i}{2}\overline{N_i\:\!^c}P_RN_i+h_{\alpha
 i}\overline{\ell_\alpha}\eta^\dag P_RN_i+\mathrm{h.c.},
\label{eq:lg}
\end{equation}
and the scalar potential $\mathcal{V}(\phi,\eta)$ is 
\begin{eqnarray}
\mathcal{V}(\phi,\eta)\!\!\!&=&
m_\phi^2\phi^\dag\phi+m_{\eta}^2\eta^\dag\eta
+\frac{\lambda_1}{2}\left(\phi^\dag\phi\right)^2
+\frac{\lambda_2}{2}\left(\eta^\dag\eta\right)^2\nonumber\\
&&\!\!\!\!\!+\lambda_3\left(\phi^\dag\phi\right)\left(\eta^\dag\eta\right)
+\lambda_4\left(\phi^\dag\eta\right)\left(\eta^\dag\phi\right)
+\frac{\lambda_5}{2}\left(\phi^\dag\eta\right)^2+\mathrm{h.c.},
\end{eqnarray}
where $\phi$ is the SM Higgs doublet.
The vacuum expectation value (VEV) of $\eta$ is assumed to be zero, so
that the discrete $\mathbb{Z}_2$ symmetry which 
guarantees the stability of DM is an exact symmetry. 
Thus Dirac neutrino masses are not generated through the
Yukawa couplings in Eq.~(\ref{eq:lg}).
After electroweak symmetry breaking, the SM Higgs $\phi$ obtains the
VEV $\left<\phi^0\right>$ and Majorana neutrino masses are generated
radiatively with the effective mass
\begin{equation}
\left(m_\nu\right)_{\alpha\beta}\simeq
\sum_{i=1}^3\frac{2\lambda_5h_{\alpha i}h_{\beta
i}\left<\phi^0\right>^2}{(4\pi)^2M_i}I\left(\frac{M_i^2}{M_\eta^2}\right),
\label{eq:neut-mass}
\end{equation}
where $M_i$ are the masses of the right-handed neutrinos $N_i$, $M_\eta^2\simeq
m_\eta^2+\left(\lambda_3+\lambda_4\right)\left<\phi^0\right>^2$,
and the loop function $I(x)$ is defined as
\begin{equation}
I\left(x\right)=\frac{x}{1-x}\left(1+\frac{x\log{x}}{1-x}\right).
\end{equation}
These relations hold for small coupling $\lambda_5$, which is needed in
order to obtain the correct neutrino masses, see below. This assumption is justified
since an extra $U(1)$ symmetry appears in the limit of $\lambda_5\to 0$.

As shown in ref.~\cite{Suematsu:2009ww}, the close to tri-bimaximal
mixing of the Pontecorvo-Maki-Nakagawa-Sakata (PMNS) mixing matrix is
achieved by adopting the following flavour structure for the Yukawa
couplings $h_{\alpha i}$ (rows are labeled by $\alpha=e,\mu,\tau$ and 
columns by $i=1,2,3$):
\begin{equation}
h_{\alpha i}=\left(
\begin{array}{ccc}
0   & 0   & h'_3\\
h_1 & h_2 & h_3\\
h_1 & h_2 & -h_3
\end{array}
\right).
\label{eq:fs}
\end{equation}
This matrix implies $\theta_{23} = \pi/4$, $\theta_{13} = 0$ and
$\tan\theta_{12} = \frac{1}{\sqrt{2}} h'_3/h_3$.\footnote{If recent
  indications~\cite{Abe:2011sj, Schwetz:2011zk, Abe:2011fz} for a non-zero value
  of the mixing angle $\theta_{13}$ should be confirmed \cite{An:2012eh}, corrections
  to Eq.~(\ref{eq:fs}) will be necessary. This will be
  discussed in the last part of Section 2.} From the current best fit
value $\sin^2\theta_{12} =
0.312^{+0.017}_{-0.015}$~\cite{Schwetz:2011zk} follows $h'_3/h_3
\approx 0.95^{+0.038}_{-0.033}$. At the
same time this Yukawa matrix allows to satisfy severe constraints from
lepton-flavour violation, see next subsection.  We write the Yukawa
couplings as $h_i=|h_i|e^{i\varphi_i}$ including the phases
$\varphi_i$. Neutrino masses are given in terms of the model
parameters as follows:
\begin{equation}
|(h_1^2+h_2^2)\Lambda_1|\simeq\frac{\sqrt{\Delta
 m_{\mathrm{atm}}^2}}{2},\quad
|h_3^2\Lambda_3|\simeq\frac{\sqrt{\Delta m_{\mathrm{sol}}^2}}{3},\quad
\mathrm{with}\quad
\Lambda_i\equiv
\frac{2\lambda_5\left<\phi^0\right>^2}{(4\pi)^2M_i}I\left(\frac{M_i^2}{M_\eta^2}\right),
\label{eq:neut}
\end{equation}
where $\Delta m_{\mathrm{atm}}^2=2.50\times10^{-3}~\mathrm{eV}^2$ and $\Delta
m_{\mathrm{sol}}^2=7.59\times10^{-5}~\mathrm{eV}^2$ correspond to the
squared-differences of the eigenvalues of the
neutrino mass matrix (\ref{eq:neut-mass}), and the mass difference of
$N_1$ and $N_2$ is neglected. The third mass eigenvalue is zero due to
the flavour structure Eq.~(\ref{eq:fs}). From Eq.~(\ref{eq:neut}) we
can estimate the required sizes for the couplings $h_{i}$ and
$\lambda_5$. Assuming $I(x) \sim 1$ we obtain
\begin{equation}\label{eq:lambda5}
\frac{\lambda_5 \, h_i^2}{10^{-11}} \sim 
\frac{M_i}{\left<\phi^0\right>}
\,
\left(\frac{\sqrt{\Delta m^2}}{0.05 \, \rm eV}\right) \,.
\end{equation}
Since $h_i$ cannot be too small because of the DM relic abundance,
typically $\lambda_5$ has to be tiny in order to obtain correct
neutrino masses. As discuss later, we impose the
perturbativity condition $|h_i|<1.5$ for the Yukawa couplings.


\subsection{Lepton flavour violation}

\begin{figure}[t]
\begin{center}
\includegraphics[scale=0.75]{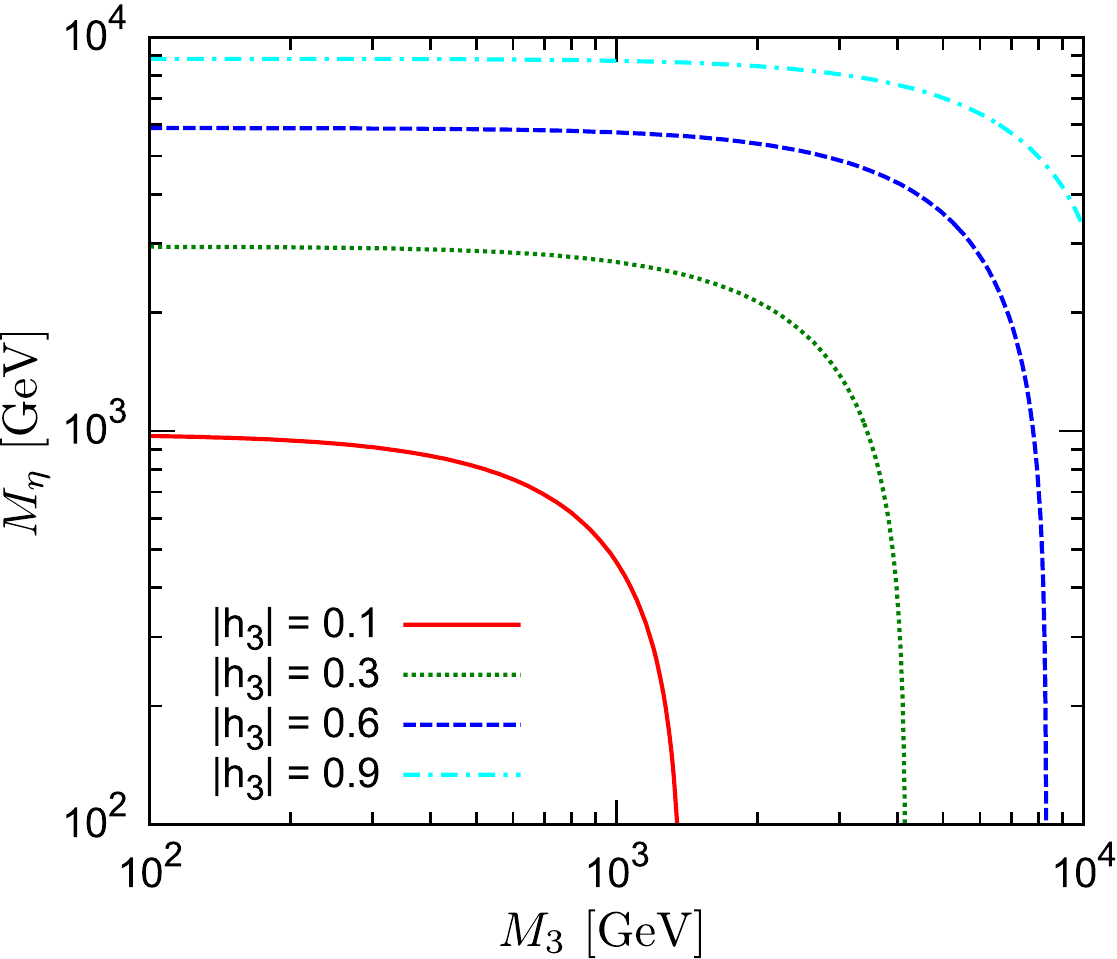}
\mycaption{Contours of $\mathrm{Br}(\mu\to e\gamma)=2.4\times
 10^{-12}$ in the ($M_3$, $M_\eta$) plane for various choices of $|h_3|$. 
 The region to the left of each contour is excluded by $\mu\to e\gamma$.}
\label{fig:lfv}
\end{center}
\end{figure}

Further constraints are imposed on the parameters by limits on charged
lepton flavour violation. The branching ratios for lepton flavour
violating processes $\ell_\alpha\to\ell_\beta\gamma$ are given as
\begin{equation}\label{eq:LFV}
\mathrm{Br}(\ell_\alpha\to\ell_\beta\gamma)=
\frac{3\alpha_{\mathrm{em}}}{64\pi
G_F^2M_\eta^4}\left|\sum_{i=1}^3h_{\alpha i}^*
h_{\beta i}F_2\left(\frac{M_i^2}{M_\eta^2}\right)\right|^2
\mathrm{Br}\left(\ell_\alpha\to\ell_\beta\nu_\alpha\overline{\nu_\beta}\right),
\end{equation}
where $\alpha_{\mathrm{em}}=e^2/(4\pi)$ is the electromagnetic fine
structure constant, $G_F$ is the Fermi constant and $M_\eta$ is the mass of
$\eta^+$ which we assume to be degenerate with $\eta^0$ for simplicity.
The function $F_2(x)$ is given by 
\begin{equation}
F_2(x)=\frac{1-6x+3x^2+2x^3-6x^2\log{x}}{6(1-x)^4}.
\end{equation}
The flavour structure of Eq.~(\ref{eq:fs}) leads to relaxed constraints
from lepton flavour violation processes such as $\mu\to e \gamma$ and
$\tau\to \mu \gamma$. Because of the two zero's in (\ref{eq:fs}) it
follows from Eq.~(\ref{eq:LFV}) that only the third right handed
neutrino mass $M_3$ and the Yukawa coupling $h_3$ contribute to
$\mu\to e\gamma$ process.
As a result, $\tau\to\mu\gamma$ gives a more
stringent constraint than $\mu\to e\gamma$ for the neutrino Yukawa
couplings $h_1$, $h_2$ and the DM mass $M_1$, and we can benefit from
the fact that the experimental upper bound
$\mathrm{Br}(\tau\to\mu\gamma)<4.5\times
10^{-8}$~\cite{Hayasaka:2007vc} is much looser than
$\mathrm{Br}(\mu\to e\gamma)<2.4\times
10^{-12}$~\cite{Adam:2011ch}.
Contours of $\mathrm{Br}(\mu\to e\gamma)=2.4\times 10^{-12}$ are shown
for several $|h_3|$ values in
Fig.~\ref{fig:lfv}.  We take $M_3=6000~\mathrm{GeV}$ and
$|h_3|=0.3$ as a benchmark point in the following discussion.  As
clear from the figure, for this choice all values of $M_\eta$ are
allowed, and for $M_\eta \lesssim 1$~TeV we predict $\mu\to e\gamma$
close to the present bound. Eq.~(\ref{eq:lambda5}) implies then
$\lambda_5 \sim 10^{-9}$. 

Thanks to the restrictions of neutrino oscillation data and
lepton-flavour violation there are very few independent parameters
left. We can choose the following set of four independent parameters:
\begin{equation}
M_\eta, \quad M_1 , \quad \delta \equiv M_2 - M_1, 
\quad \xi \equiv {\rm Im}(h_2^*h_1) \,,
\end{equation}
with $\delta \ll M_1$. Since we fix $h_3$ and $M_3$ to the benchmark
point above in order to satisfy $\mu\to e\gamma$, the relations
Eq.~(\ref{eq:neut}) determine $\lambda_5$ as well as $|(h_1^2+h_2^2)|$
for a given choice of $M_\eta$ and $M_1$. However, there is still an
undetermined relative phase between $h_1$ and $h_2$, and we define the
parameter $\xi$, which will play an important role in the following.

\subsection{DM relic abundance}

We assume that the lightest right handed neutrino $N_1$ is the lightest of
the $\mathbb{Z}_2$-odd particles, and hence it will be stable and serve as
the DM candidate. We assume that it is almost degenerated with the second
lightest right handed neutrino $N_2$. The mass degeneracy could
be provided by imposing a symmetry for the right handed neutrinos $N_i$.
This could be for example the conservation of particle number such that
$N_1$ and $N_2$ form a pseudo-Dirac particle \cite{TuckerSmith:2001hy}. The
smallness of the mass splitting is then related to suppressed operators
violating the symmetry. The $N_i$ couple to the SM only via the Yukawa
interaction with the lepton doublet and therefore our DM is
leptophilic.\footnote{Another motivation for leptophilic DM may come from
cosmic ray observations from the PAMELA~\cite{Adriani:2008zr} and
Fermi-LAT~\cite{Ackermann:2010ij, Ackermann:2011rq} experiments, finding an
excess of positrons but anti-protons in agreement with expectations. In
order to obtain the required count rates, however, the annihilation cross
section must be boosted by a mechanism such as
Sommerfeld~\cite{Hisano:2004ds} or Breit-Wigner
enhancement~\cite{Feldman:2008xs, Ibe:2008ye}, beyond the model considered
here.} The relic density and indirect detection of DM in the model have
been investigated with the flavour structure of Eq.~(\ref{eq:fs}) in
refs.~\cite{Suematsu:2009ww, Suematsu:2010gv, Suematsu:2010nd,
Fukuoka:2010kx}.  Here we investigate the prospects for direct detection of
DM in this setup for the first time.

For the thermal production of DM in this model co-annihilations between
$N_1$ and $N_2$ have to be considered, since they are assumed to be
highly degenerate, leading to an enhanced effective annihilation cross
section~\cite{Griest:1990kh}.  The effective annihilation cross
section is written as
$\sigma_{\mathrm{eff}}v=a_{\mathrm{eff}}+b_{\mathrm{eff}}v^2+\mathcal{O}(v^4)$.
Then the approximate analytic solution of the Boltzmann equation which
describes the evolution of the DM density is given by
\begin{equation}
\Omega h^2\simeq
\frac{1.07\times 10^{9}x_f~[\mathrm{GeV^{-1}}]}
{\sqrt{g_*}m_{\mathrm{pl}}\left(a_{\mathrm{eff}}+3b_{\mathrm{eff}}/x_f\right)},
\quad
\mbox{with}
\quad
x_f=\frac{M_1}{T_f},
\end{equation}
where $g_{*}$ is the number of relativistic degrees of freedom at the time
of freeze-out $T_f$ and $m_{\mathrm{pl}}=1.2\times 10^{19}~\mathrm{GeV}$. 
WMAP data~\cite{Komatsu:2010fb} implies $\Omega h^2=0.11260\pm0.0036$.
Taking into account co-annihilations of $N_1$ and $N_2$ we find for the
coefficients $a_{\mathrm{eff}}$ and $b_{\mathrm{eff}}$ in the effective 
annihilation cross section
\begin{eqnarray}
a_{\mathrm{eff}}\!\!\!&=&\!\!\!\frac{\xi^2}{2\pi}
\frac{M_1^2}{\left(M_\eta^2+M_1^2\right)^2} \,,\label{eq:aeff}\\
b_{\mathrm{eff}}\!\!\!&=&\!\!\!
\frac{|h_1^2+h_2^2|^2}{24\pi}
\frac{M_1^2\left(M_\eta^4+M_1^4\right)}{\left(M_\eta^2+M_1^2\right)^4}
+\frac{\xi^2}{2\pi}
\frac{M_1^2\left(M_\eta^4-3M_\eta^2M_1^2-M_1^4\right)}{\left(M_\eta^2+M_1^2\right)^4} \,,
\label{eq:beff}
\end{eqnarray}
where the effect of the mass difference between $N_1$ and $N_2$ is
assumed to be negligible. The terms proportional to $\xi^2$ come from
the co-annihilation process
$N_1N_2\to\ell_\alpha\overline{\ell_\beta}$, whereas the $N_1N_1$ and
$N_2N_2$ annihilations lead to the terms proportional to $h_1^2$ and
$h_2^2$, respectively. We observe from Eq.~(\ref{eq:aeff}) and
(\ref{eq:beff}) that the $s$-wave ($a_{\mathrm{eff}}$-term) is only present due to
co-annihilations. If there is no phase difference between $h_1$ and
$h_2$, the combination of the neutrino Yukawa couplings $\xi$ vanishes
and only $p$-wave annihilation remains. This corresponds to
the helicity suppression for a Majorana fermion. Thus co-annihilations and a
non-zero phase difference play an important role in obtaining the
correct DM relic density.

\begin{figure}[t]
\begin{center}
\includegraphics[scale=0.75]{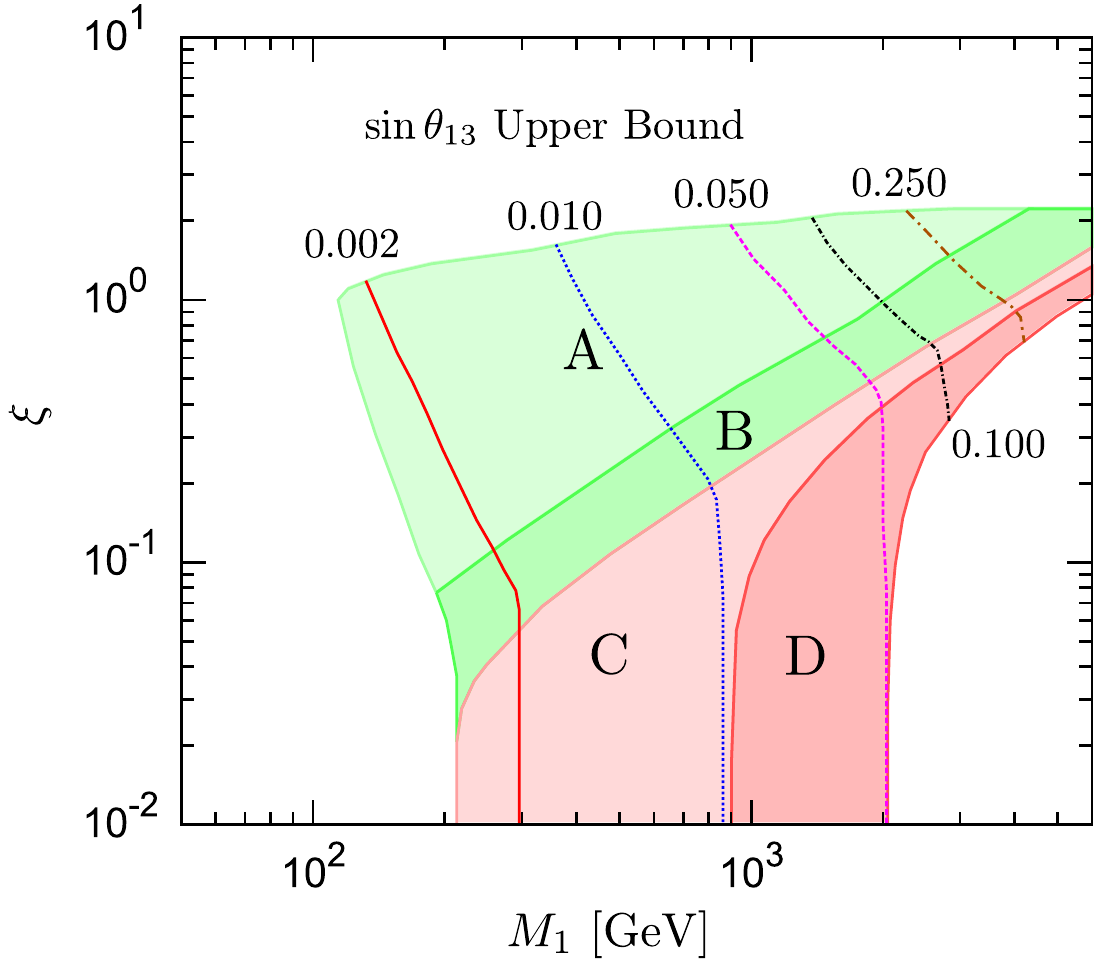} 

\mycaption{Region in the space of DM mass $M_1$ and
  $\xi = |h_1h_2|\sin(\varphi_1 - \varphi_2)$ consistent with
  neutrino masses and mixing, lepton flavour violation,
  perturbativity, and the relic density of DM. The regions with
  different color shadings denoted by A, B, C, D, correspond to
  different assumptions on $M_\eta$, with A: $2.0 < M_\eta/M_1 <
  {9.8}$, B: $1.2 < M_\eta/M_1 < 2.0$, C: $1.05 < M_\eta/M_1 < 1.20$,
  D: $1.0 < M_\eta/M_1 < 1.05$. The curves show the
  upper bound on $\sin\theta_{13}$ from $\mu\to e\gamma$
  when the Yukawa matrix Eq.~(\ref{eq:fs})
  is extended to Eq.~(\ref{eq:ykw-ex}).}
\label{fig:relic}
\end{center}
\end{figure}

For the following results we use the micrOMEGAs
package~\cite{Belanger:2008sj} to calculate numerically the relic
abundance of DM. In addition to $N_1-N_2$ co-annihilations, also
co-annihilations with $\eta$ are important, if $M_\eta$ becomes close
to $M_1$. The allowed parameter region in the plane of DM mass and the
Yukawa coupling $\xi$, which is consistent with neutrino masses and
mixings, lepton flavour violation, and DM relic density is shown in
Fig.~\ref{fig:relic}. The allowed region is colored and divided into
four regions A, B, C, D, corresponding to different assumptions on the
ratio $M_1/M_\eta$. The upper bound on $\xi$ is imposed by requiring
perturbativity of the Yukawa couplings. The lower bound on $M_1$ in
regions A and B is determined by the limit on $\tau\to\mu\gamma$
together with the relic abundance requirement. There is no allowed
parameter region if $M_\eta/M_1 \gtrsim 9.8$ because taking into
account perturbativity as well as $\tau\to\mu\gamma$ the annihilation
cross section is suppressed by $M_\eta^4$. If $M_\eta/M_1$ comes close
to 9.8 we are driven to the left-upper corner of the allowed region in
Fig.~\ref{fig:relic}. In the parameter region C and D we have
$M_\eta/M_1 < 1.2$ and co-annihilations with $\eta$ become
important. Without co-annihilations with $\eta$, the parameter space C
and D would not appear, and we would obtain a lower bound on
$|\xi|$. However, if $N_1-\eta$ co-annihilations are relevant the
correct relic density can be obtained even for vanishing $\xi$. In all
cases we can conclude that the correct relic density is always 
obtained thanks to co-annihilations with either $N_2$ or
$\eta$. 


If $M_\eta \approx M_1$ one may worry about a long-lived charged
particle contained in the doublet $\eta=(\eta^+,\eta^0)$. For
instance, the predictions for Big Bang Nucleosynthesis (BBN) may be
altered by the energy injection due to the decay of $\eta^+$ into
charged leptons~\cite{Kawasaki:2004qu, Jedamzik:2006xz}. 
We have checked that for the
parameter ranges of interest $\eta$ decays much faster than 0.01~s
unless it is degenerate with $N_1$ at the level of $10^{-10}$, and
hence BBN will be not affected.

Let us now consider deviations from the flavour structure assumed
in Eq.~(\ref{eq:fs}). This will become necessary if hints for a non-zero
$\theta_{13}$~\cite{Abe:2011sj, Abe:2011fz} should be confirmed \cite{An:2012eh}.  In this
case we expect additional constraints from $\mu\to e\gamma$. Let us consider
a small perturbation of Eq.~(\ref{eq:fs}) as
\begin{equation}
h_{\alpha i}=\left(
\begin{array}{ccc}
\epsilon_1 & \epsilon_2 & h_3'\\
h_1 & h_2 & h_3\\
h_1 & h_2 & -h_3
\end{array}
\right)+\mathcal{O}(\epsilon^2),
\label{eq:ykw-ex}
\end{equation}
with $\epsilon_i \ll h_j$. Then we allow a non-zero value 
$\sin\theta_{13}=\epsilon_3$ and deviations of $\theta_{23}$ from $\pi/4$ as
$\sin\theta_{23}=1/\sqrt{2}+\epsilon_{4}$, with $\epsilon_{3,4} \ll 1$.
Diagonalizing the neutrino mass matirx Eq.~(\ref{eq:neut-mass}) we obtain
at linear order in $\epsilon_i$
\begin{eqnarray}
\epsilon_4
\!\!\!&=&\!\!\!
\frac{1}{\sqrt{2}}\frac{\tan\theta_{12}h_3^2\Lambda_3}
{\left(h_1^2+h_2^2\right)\Lambda_1-h_3^2\Lambda_3}\epsilon_3,\\
\epsilon_1h_1+\epsilon_2h_2
\!\!\!&=&\!\!\!
\sqrt{2}\left(h_1^2+h_2^2\right)
\frac{\left(h_1^2+h_2^2\right)\Lambda_1-\sec^2\theta_{12}h_3^2\Lambda_3}
{\left(h_1^2+h_2^2\right)\Lambda_1-h_3^2\Lambda_3}\epsilon_3\equiv P\epsilon_3.
\end{eqnarray}
If we assume that $\epsilon_1$, $\epsilon_2$ and $\epsilon_3$ are real, we
obtain from Eq.~(\ref{eq:LFV}) the following expression for 
$\mu\to e\gamma$:
\begin{equation}\label{eq:m2eg}
\mathrm{Br}\left(\mu\to e\gamma\right)=
\frac{3\alpha_{\mathrm{em}}}{64\pi G_F^2M_\eta^4}
\left|
P\epsilon_3F_2\left(\frac{M_1^2}{M_\eta^2}\right)
+\sqrt{2}\tan\theta_{12}|h_3|^2F_2\left(\frac{M_3^2}{M_\eta^2}\right)
\right|^2.
\end{equation}
Thus a non-zero $\theta_{13}$ directly gives a contribution to $\mu\to
e\gamma$ and $\epsilon_3=\sin\theta_{13}$ is constrained by the limit
on this process. Using Eq.~(\ref{eq:neut}) the parameter $P$ is
approximately obtained as $P\approx \sqrt{2}\left(h_1^2+h_2^2\right)$,
and we can obtain an upper bound on $\epsilon_3$ from $\mu\to e
\gamma$ at each point in Fig.~\ref{fig:relic}.\footnote{In general,
  the phase of $P$ depends on the phases of the Yukawa couplings $h_1$
  and $h_2$, i.e., $\varphi_1$ and $\varphi_2$, not only the phase
  difference $\varphi_1-\varphi_2$. For simplicity we set the overall
  phase of $P$ to zero. This phase might play a role if the two terms
  in Eq.~(\ref{eq:m2eg}) are of comparable size.}  Contours of the
upper bound on $\sin\theta_{13}$ are shown in
Fig.~\ref{fig:relic}. The upper bound becomes severe for small DM
mass. Recent results of a non-zero $\theta_{13}$~\cite{An:2012eh}
imply $\sin\theta_{13} > 0.1$ at $3\sigma$. According to
Fig.~\ref{fig:relic} this requires DM masses around the TeV scale with
$\xi\sim\mathcal{O}(0.1-1)$.

In addition to the extension of the Yukawa matrix
Eq.~(\ref{eq:ykw-ex}) we checked also the effect of changing the
$\tau1$ and $\tau2$ components into $h_1+\epsilon_1$ and
$h_2+\epsilon_2$. The factor $P$ only changes to $P\approx
13\sqrt{2}(h_1^2+h_2^2)/6$. Moreover, changing the $\tau 3$ component
of Eq.~(\ref{eq:ykw-ex}) into $-(h_3+\epsilon)$, the deviation
$\epsilon$ is required to be zero up to $\mathcal{O}(\epsilon)$ from
the diagonalization condition of the neutrino mass matrix. Therefore
these extensions of the Yukawa matrix do not change the analysis
drastically.

\section{Direct Detection of Leptophilic DM}
\label{sec:dd}

\subsection{Inelastic Scattering Cross Section}

\begin{figure}[t]
\begin{center}
\includegraphics[scale=.9]{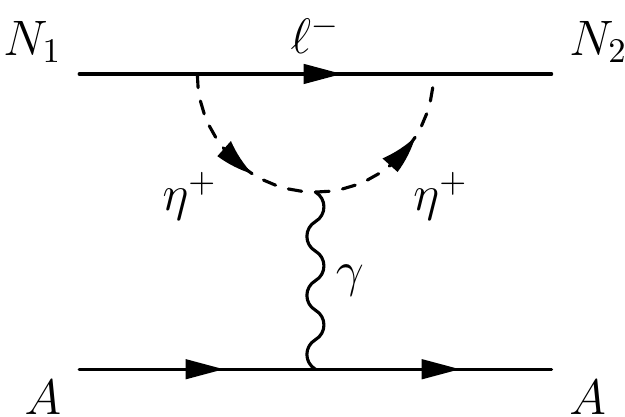}
\qquad
\includegraphics[scale=.9]{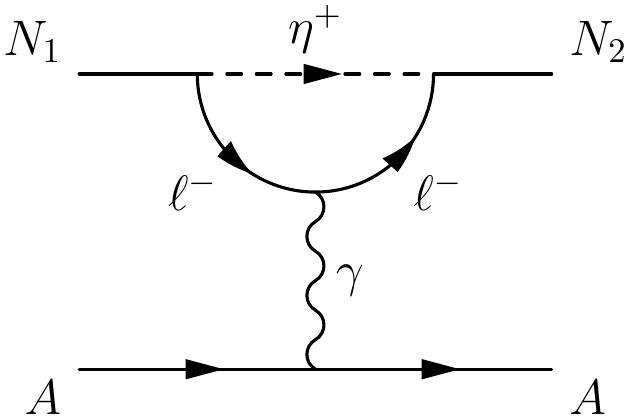}
\mycaption{Diagrams for the inelastic scattering process of $N_1$ off 
 a nucleus $A$.}
\label{fig:dd-inel}
\end{center}
\end{figure}

Inelastic scattering occurs
through the effective interactions with quarks which come from the
1-loop diagrams shown in Fig.~\ref{fig:dd-inel}.
The 3-point vertex effective interactions of $N_1$, $N_2$ and $\gamma$
which give a dominant contribution to the inelastic scattering are written as
\begin{equation}
\mathcal{L}_{\mathrm{eff}}=
ia_{12}\overline{N_2}\gamma^{\mu}N_1\partial^{\nu}F_{\mu\nu}
+i\left(\frac{\mu_{12}}{2}\right)\overline{N_2}\sigma^{\mu\nu}N_1F_{\mu\nu}
+ic_{12}\overline{N_2}\gamma^{\mu}N_1A_{\mu},
\label{eq:eff}
\end{equation}
where the factor $i$ is a conventional factor to obtain real couplings
$a_{12}$, $c_{12}$ and $\mu_{12}$, and $F_{\mu\nu}$ is the electromagnetic
field strength.  The coefficient $\mu_{12}$ is known as the transition
magnetic moment between $N_1$ and $N_2$. Elastic scattering does not occur
through the effective interactions because the operators
$\overline{N_1}\gamma^\mu N_1$ and $\overline{N_1}\sigma^{\mu\nu}N_1$ are
identical zero for Majorana fermions. General inelastic scattering of DM has
been discussed in refs.~\cite{TuckerSmith:2001hy, TuckerSmith:2004jv}, and
inelastic scattering due to the magnetic moment interactions in
ref.~\cite{Chang:2010en}. Loop induced DM--nucleus scattering for
leptophilic DM has been pointed out in ref.~\cite{Kopp:2009et}, and the model
considered here is a specific realization of ``flavoured'' DM discussed in
ref.~\cite{Agrawal:2011ze}, where similar diagrams to the ones from
Fig.~\ref{fig:dd-inel} have been considered. For another recent model for
magnetic inelastic DM see ref.~\cite{patra}.

In the model considered here, the coefficients $a_{12}$, $c_{12}$ and $\mu_{12}$ are calculated as 
\begin{eqnarray}
a_{12}\!\!\!&=&\!\!\!-\sum_{\alpha}
\frac{\mathrm{Im}\left(h_{\alpha 2}^*h_{\alpha 1}\right)e}
{2(4\pi)^2M_\eta^2}I_{\mathrm{a}}\left(\frac{M_1^2}{M_\eta^2},\frac{m_\alpha^2}{M_\eta^2}\right),
\label{eq:a12}\\ 
\mu_{12}\!\!\!&=&\!\!\!-\sum_\alpha
\frac{\mathrm{Im}\left(h_{\alpha 2}^*h_{\alpha 1}\right)e}{2(4\pi)^2M_\eta^2}
2M_1I_\mathrm{m}\left(\frac{M_1^2}{M_\eta^2},\frac{m_{\alpha}^2}{M_\eta^2}\right),
\label{eq:mu12}\\
c_{12}\!\!\!&=&\:\:\sum_\alpha
\frac{\mathrm{Im}\left(h_{\alpha 2}^*h_{\alpha 1}\right)e}{2(4\pi)^2M_\eta^2}
q^2I_\mathrm{c}\left(\frac{M_1^2}{M_\eta^2},\frac{m_{\alpha}^2}{M_\eta^2}\right),
\label{eq:c12}
\end{eqnarray}
where $q^2$ is the momentum transfer and the explicit forms of the function 
$I_\mathrm{a}(x,y)$, $I_{\mathrm{m}}(x,y)$ and $I_{\mathrm{c}}(x,y)$, which
come from the loop integrals, are given in Appendix~A.  Eq.~(\ref{eq:fs})
implies that $\mathrm{Im}\left(h_{\alpha 2}^*h_{\alpha 1}\right) = \xi$ and
therefore the parameter $\xi$ responsible for $N_1-N_2$ co-annihilations
controls also the effective interactions of DM with nuclei. 

From the effective interactions, we can obtain three types of
differential scattering cross sections with a nucleus which has atomic
number $Z$, mass number $A$, mass $m_A$, spin $J_A$ and magnetic moment $\mu_A$, see
e.g., \cite{Chang:2010en, Agrawal:2011ze}:
\begin{eqnarray}
\frac{d\sigma_{\mathrm{CC}}}{dE_R}\!\!\!&=&\!\!\!\frac{Z^2b_{12}^2m_A}{2{\pi}v^2}F^2(E_R),
\label{eq:zz}\\
\frac{d\sigma_{\mathrm{DC}}}{dE_R}\!\!\!&=&\!\!\!\frac{Z^2\alpha_{\mathrm{em}}\mu_{12}^2}{E_R}
\left[1-\frac{E_R}{v^2}\left(\frac{1}{2m_A}+\frac{1}{M_1}\right)
-\frac{\delta}{v^2}\frac{1}{\mu_{\mathrm{DM}}}
-\frac{\delta^2}{v^2}\frac{1}{2m_AE_R}\right]F^2(E_R),
\label{eq:dz}\\
\frac{d\sigma_{\mathrm{DD}}}{dE_R}\!\!\!&=&\!\!\!\frac{\mu_A^2\mu_{12}^2m_A}
{{\pi}v^2}\left(\frac{J_A+1}{3J_A}\right)
F_D^2(E_R),
\label{eq:dd}
\end{eqnarray}
with the coefficient 
\begin{equation}\label{eq:b12}
b_{12}=(a_{12}+c_{12}/q^2)e \,.
\end{equation}
The cross sections Eq.~(\ref{eq:zz}),~(\ref{eq:dz})~and~(\ref{eq:dd})
are called charge-charge (CC), dipole-charge (DC), and dipole-dipole
(DD) coupling, respectively. Here $E_R$ is the recoil energy, the
parameter $\delta$ is the mass difference between $N_2$ and $N_1$
i.e., $\delta=M_2-M_1$ and $\mu_{\mathrm{DM}}$ is the DM--nucleus
reduced mass.  Magnetic moments of several nuclei are shown in
Tab.~\ref{tab:mag}.  $F(E_R)$ is the nuclear form factor for which we
use the parametrization
\begin{equation}
F(E_R)=
\frac{3\left[
\sin({\kappa}r)-{\kappa}r\cos({\kappa}r)\right]}
{({\kappa}r)^3}e^{-\kappa^2s^2/2},
\end{equation}
with $\kappa=\sqrt{2m_AE_R}$, $r=\sqrt{R^2-5s^2}$, $R\simeq
1.2A^{1/3}~\mathrm{fm}$ and $s\simeq 1~\mathrm{fm}$. 
$F_D(E_R)$ is the nuclear magnetic form factor and it is not well-known, see
e.g., the discussion in ref.~\cite{Chang:2010en}.
We adopt the following approximation for $F_D(E_R)$. The magnetic moment of
a nucleus receives contributions from the spin $\langle S_{n,p}\rangle$ as
well as orbital momentum $\langle L_{n,p}\rangle$
of the neutrons and protons: 
\begin{equation}\label{eq:mu}
\mu_A=
g_p^s \left<S_p\right> + g_n^s\left<S_n\right> + 
g_p^l \left<L_p\right> + g_n^l\left<L_n\right>.
\end{equation}
We approximate the magnetic form factor by neglecting the orbital
momentum contribution\footnote{The ratio of spin and orbital
  contributions to the magnetic moment in Eq.~(\ref{eq:mu}) are $0.59
  : 0.41$ for Sodium, $0.52 : 0.48$ for Iodine, $0.96 : 0.04$ for
  Xenon, $-0.38 : 1.38$ for Cesium. Therefore, neglecting the orbital
  contribution is an excellent approximation for Xenon. For the other
  nuclei this introduces an error of about an factor 2 and therefore
  the limits derived from KIMS and DAMA should be considered only
  approximate.} and use the spin from factors weighted by the
corresponding $g^s$ factors:
\begin{equation}
F_D(E_R) \approx \frac{g_p^s S_p(q^2) + g_n^s S_n(q^2)}
{g_p^s S_p(0) + g_n^s S_n(0)} \,.
\end{equation}
The spin-dependent form factors and $g^s_{p,n}$ factors are taken from
refs.~\cite{Ressell:1997kx, Bednyakov:2006ux}. 

\begin{table}[t]
\begin{center}
\begin{tabular}{|c||c|c|c|c|c|c|c|}\hline
 & $^{19}_{\:\:9}$F &$^{23}_{11}$Na & $^{73}_{32}$Ge & $^{127}_{\:\:53}$I & $^{131}_{\:\:54}$Xe 
& $^{133}_{\:\:55}$Cs & $^{183}_{\:\:74}$W\\\hhline{|=#=|=|=|=|=|=|=|}
$J_A$ & $1/2$ & $3/2$ & $9/2$ & $5/2$ & $3/2$ & $7/2$ & $1/2$\\\hline
$\mu_A/\mu_{N}$ & $2.629$ & $2.218$ & $-0.879$ & $2.813$ &
		     $0.692$ & $2.582$ & $0.118$\\\hline
\end{tabular}
\mycaption{Magnetic moments for several nuclei in units of $\mu_{N}$
 where $\mu_{N}=e/2m_p$ is the nuclear
 magneton~\cite{Ellis:1991ef}.}
\label{tab:mag}
\end{center}
\end{table}

In addition to the CC, DC, DD interactions from Eqs.~(\ref{eq:zz}),
(\ref{eq:dz}), (\ref{eq:dd}) also a charge-dipole coupling exists.
However there is an additional suppression factor of $q^2$ compared to
the other couplings, thus it can be neglected.  The DC
coupling is singular at $E_R=0$.  Therefore the predicted event rate
of the DC coupling is enhanced at low recoil energies due to the
singularity, and we cannot define a total cross section at the zero
momentum transfer limit $\sigma_{\mathrm{DC}}^0$. This situation is the same as
in Coulomb scattering.


\subsection{Comparison of the Predicted Event Rate with Experiments}

We compare the event rate calculated from the effective interactions with
XENON100\cite{Aprile:2011hi}, KIMS~\cite{kims-taup} and
DAMA~\cite{Bernabei:2010mq} data.  The DD coupling might be
important for KIMS or DAMA~\cite{Chang:2010en} since in these experiments, 
the target nuclei are iodine (I) and cesium (Cs) for KIMS, iodine and sodium
(Na) for DAMA, which have a large nuclear magnetic moment as can be seen
from Tab.~\ref{tab:mag}.  The event rate is written as
\begin{equation}
\frac{dR}{dE_R}=\sum_{\mathrm{nuclei}}\frac{\rho_\odot}{M_1}\frac{1}{M_\mathrm{det}}
\int_{v>v_{\mathrm{min}}}\frac{d\sigma}{dE_R}vf(\bm{v})d^3v,
\end{equation}
where $\rho_{\odot}\simeq 0.3~\mathrm{GeVcm^{-3}}$ is the local DM
density, $M_{\mathrm{det}}$ is the mass of target material,
$v_{\mathrm{min}}$ is the minimum velocity 
required for DM to scatter off a nucleus with recoil energy $E_R$,
\begin{equation}
v_{\mathrm{min}}=\frac{1}{\sqrt{2m_AE_R}}\left(\frac{m_AE_R}{\mu_{\mathrm{DM}}}+\delta\right),
\end{equation}
and $f(\bm{v})$ is the local DM velocity distribution function in
the rest frame of the Earth.
It is obtained by a Galilean transformation from a
Maxwell-Boltzmann distribution in the rest frame of the galaxy with the
velocity dispersion $v_0=220~\mathrm{km/s}$ and the escape velocity from
the galaxy $v_{\mathrm{esc}}=544~\mathrm{km/s}$. 
The velocity distribution function $f(\bm{v})$ is normalized to
$\int f(\bm{v})d^3v=1$. 
The relative velocity of the Earth to the galaxy is
$v_e=v_\odot+v_{\mathrm{orb}}\cos\gamma\cos\left[2\pi(t-t_0)/\mathrm{year}\right]$
with $v_\odot=v_0+12~\mathrm{km/s}$, $v_{\mathrm{orb}}=30~\mathrm{km/s}$,
$\cos\gamma=0.51$ and $t_0=$ June 2nd. 
We must evaluate the following velocity integrals to predict the 
event rate:
\begin{eqnarray}
\zeta_1(v_{\mathrm{min}},v_e)\!\!\!&=&\!\!\!\int_{v_{\mathrm{min}}}^\infty
\frac{f(\bm{v}+\bm{v}_e)}{v}d^3v,\\
\zeta_2(v_{\mathrm{min}},v_e)\!\!\!&=&\!\!\!\int_{v_{\mathrm{min}}}^\infty
vf(\bm{v}+\bm{v}_e)d^3v.
\end{eqnarray}
The analytic formulas for the DM velocity integrals given in
refs.~\cite{McCabe:2010zh, Fitzpatrick:2010br} are used.  The total
predicted event rate in the XENON100, DAMA, and KIMS experiments is
obtained by integrating the differential event rate with respect to an
appropriate recoil energy range.  We use the energy range and the
quenching factors shown in Tab.~\ref{tab:qf}.  The quenching factor
is the ratio of the energy deposited in scintillation
light to the total nuclear recoil energy.

\begin{table}[t]
\begin{center}
\begin{tabular}{|c||c|c|}\hline
         & Energy range  & Quenching factor   \\\hhline{|=#=|=|}
XENON100 & $8.4 - 44.6$ keV  & $-$              \\\hline
KIMS     & $3.6 - 5.8$ keVee & 0.1 (Cs), 0.1 (I)  \\\hline
DAMA     & $2 - 8$ keVee     & 0.3 (Na), 0.09 (I) \\\hline
\end{tabular}
\mycaption{The energy range and the quenching factor for the
  experiments XENON100~\cite{Aprile:2011hi}, KIMS~\cite{kims-taup},
  and DAMA~\cite{Bernabei:2010mq}.  For XENON100 we use the same
  light-yield function $L_\mathrm{eff}$ as in
  ref.~\cite{Aprile:2011hi}.}
\label{tab:qf}
\end{center}
\end{table}

\begin{figure}[t]
\begin{center}
\includegraphics[scale=0.6]{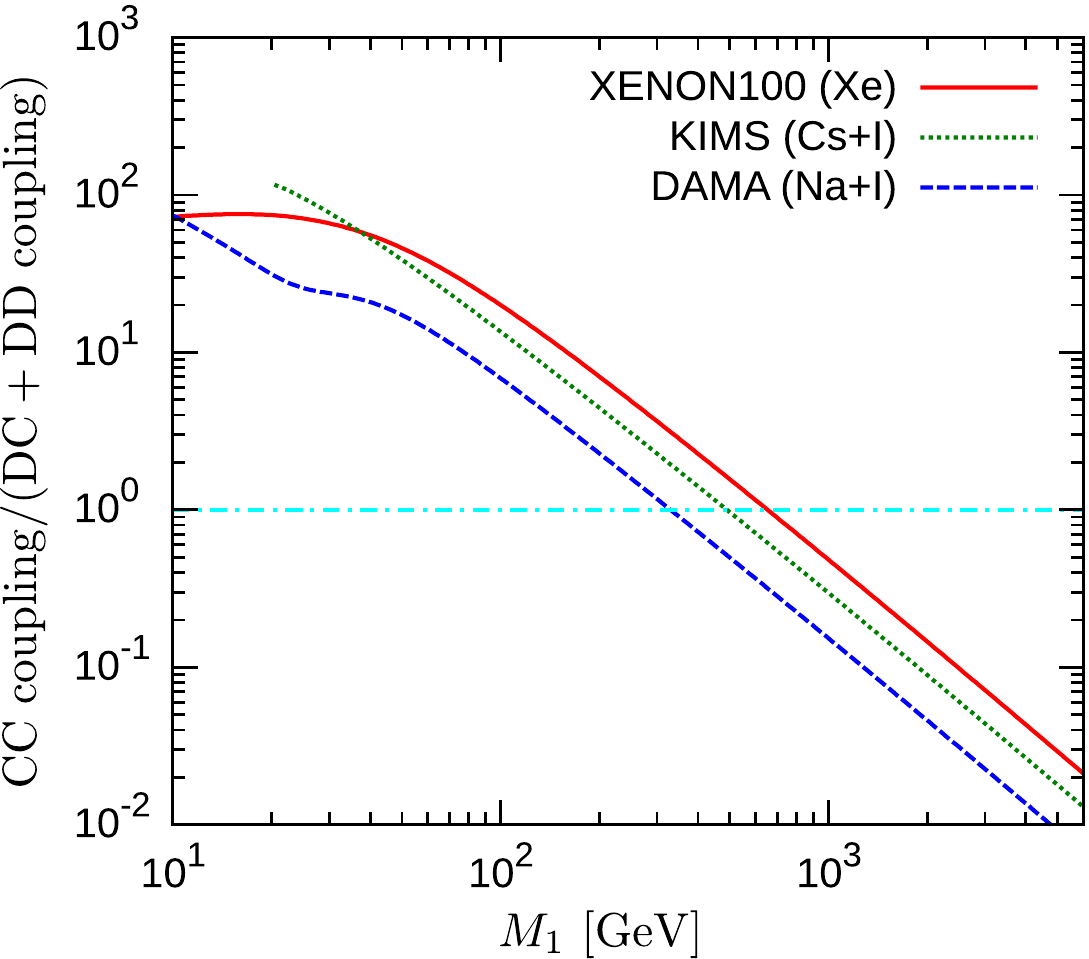} \qquad
\includegraphics[scale=0.6]{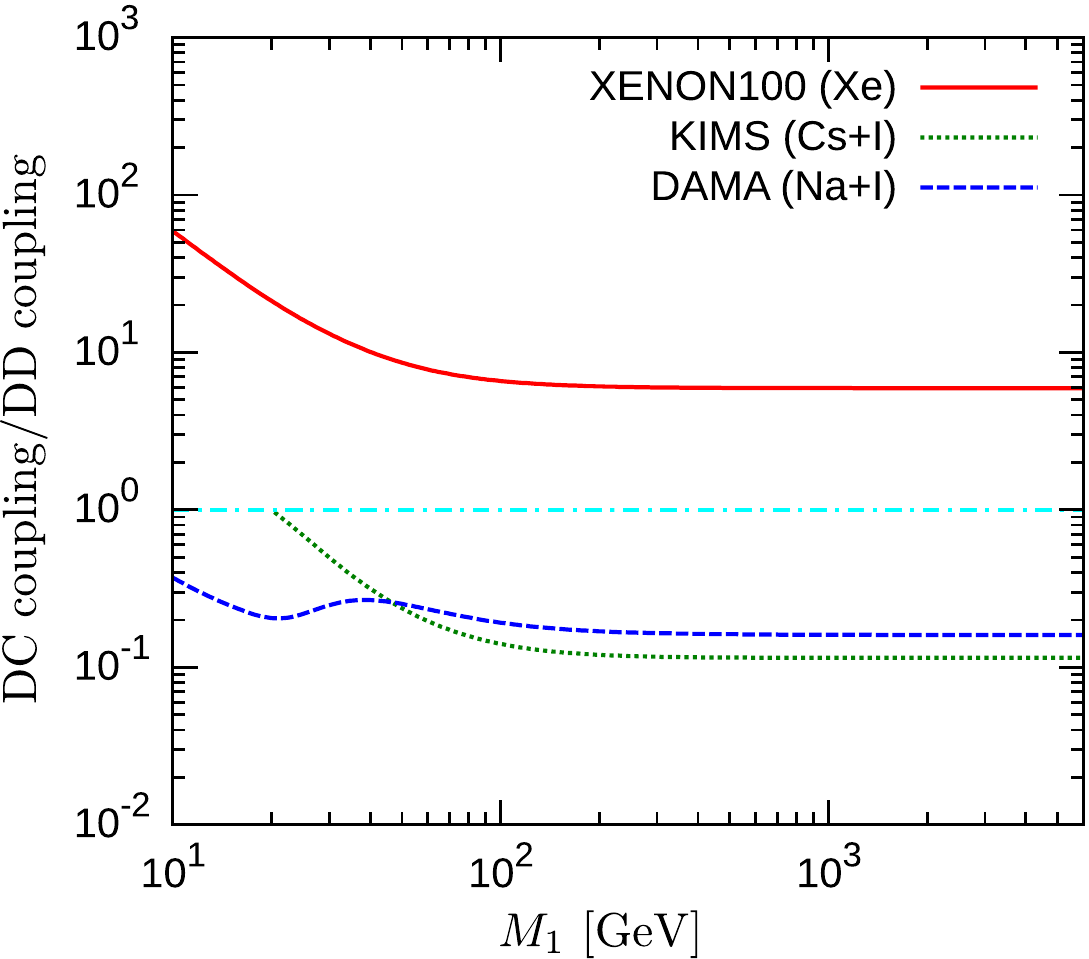} 
  
  \mycaption{Relative contributions of the charge-charge (CC),
    dipole-charge (DC), and dipole-dipole (DD) interactions to the
    total predicted event rate in XENON100, KIMS, and DAMA. The left
    panel shows the contribution from CC relative to the sum of DC and
    DD, the right panel shows the ratio of the DC and DD
    contributions. We assume $M_\eta / M_1 = 1.5$ and $\delta =
    0$.}
\label{fig:ratio}
\end{center}
\end{figure}

In Fig.~\ref{fig:ratio} we illustrate the relative importance of the
the CC, DC, DD interactions from Eqs.~(\ref{eq:zz}), (\ref{eq:dz}),
(\ref{eq:dd}) for the XENON100, KIMS, and DAMA experiments by
calculating the total event rate induced from each of the three
interaction types separately. We observe from the left panel that
typically CC interactions are more important for small masses $M_1$,
which follows from the different dependence on the DM mass of $b_{12}$
and $\mu_{12}$. The value $M_1$ where CC becomes subdominant depends on
the ratio $M_\eta/M_1$. The right panel of Fig.~\ref{fig:ratio} shows
that for XENON100 the DC coupling is more important, whereas for KIMS and
DAMA DD dominates, because of the large magnetic moments of iodine and
sodium. The features of the DAMA curves around $M_1 \simeq 20$~GeV in
both panels are a consequence of the presence of the two elements (I
and Na) with rather different masses. 
In general the relative importance of CC, DC, DD depends on
the ratio $M_\eta/M_1$ and to a lesser extent on $\delta$. The main
conclusion is that depending on the region in the parameter space and
depending on the considered experiment, any of the three interaction
types can be important and all of them have to be taken into account.

\begin{figure}[t!]
\begin{center}
\includegraphics[scale=0.65]{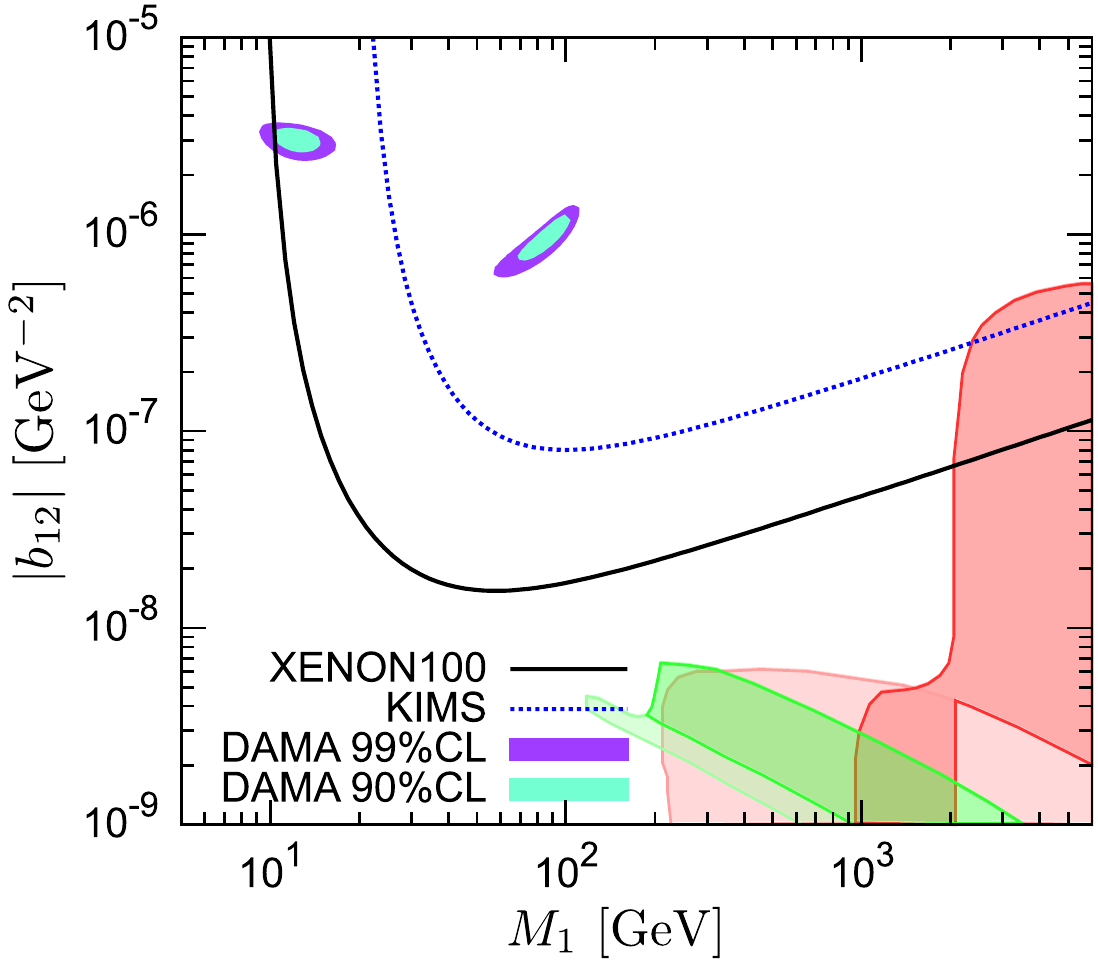} \qquad
\includegraphics[scale=0.65]{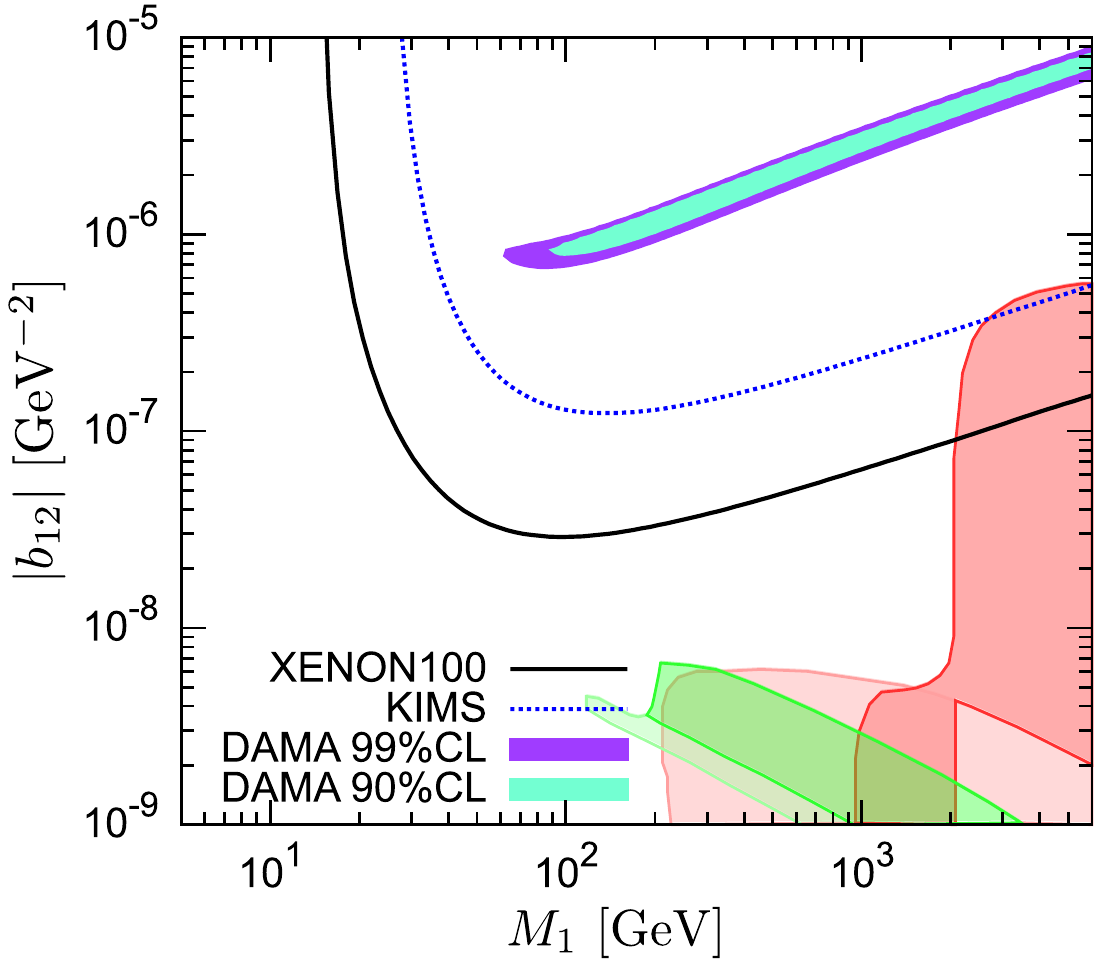} \\\quad\\ 
\includegraphics[scale=0.65]{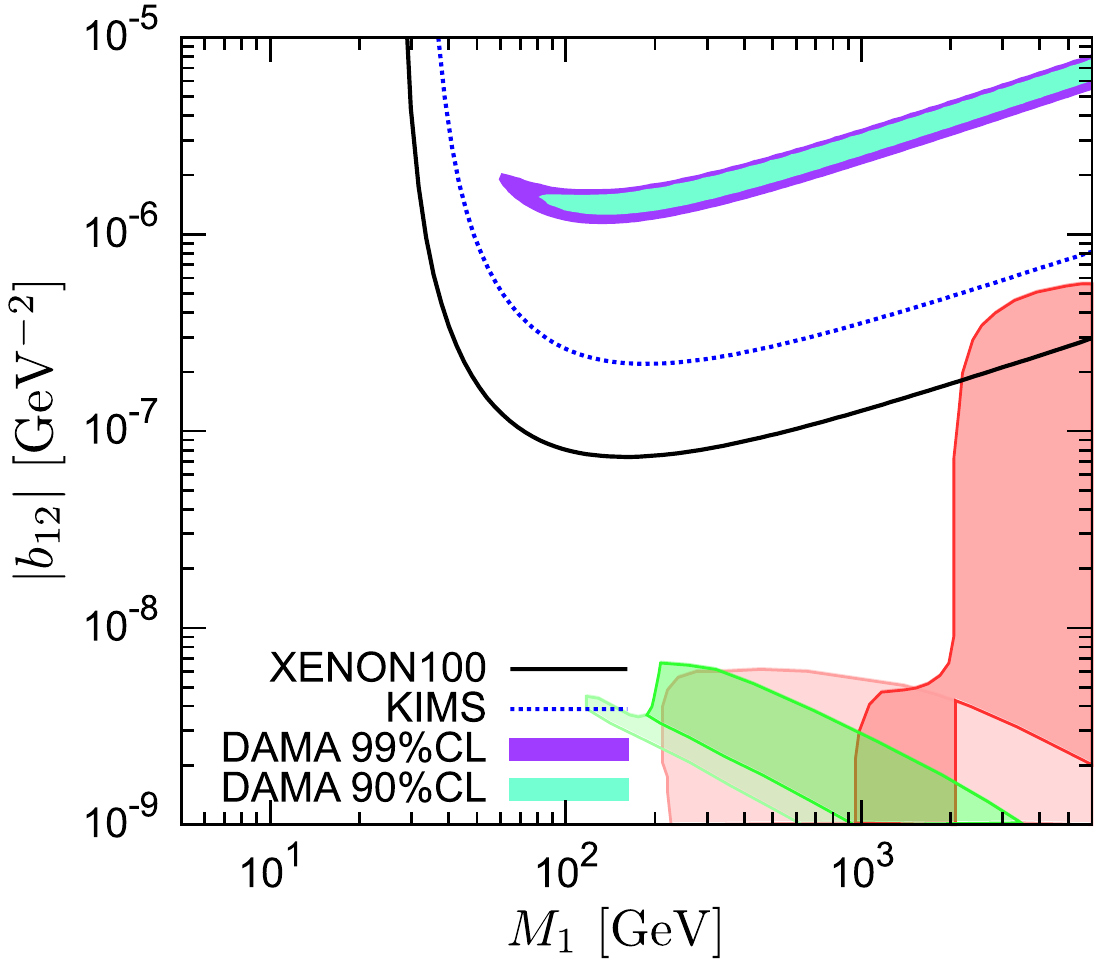} \qquad
\includegraphics[scale=0.65]{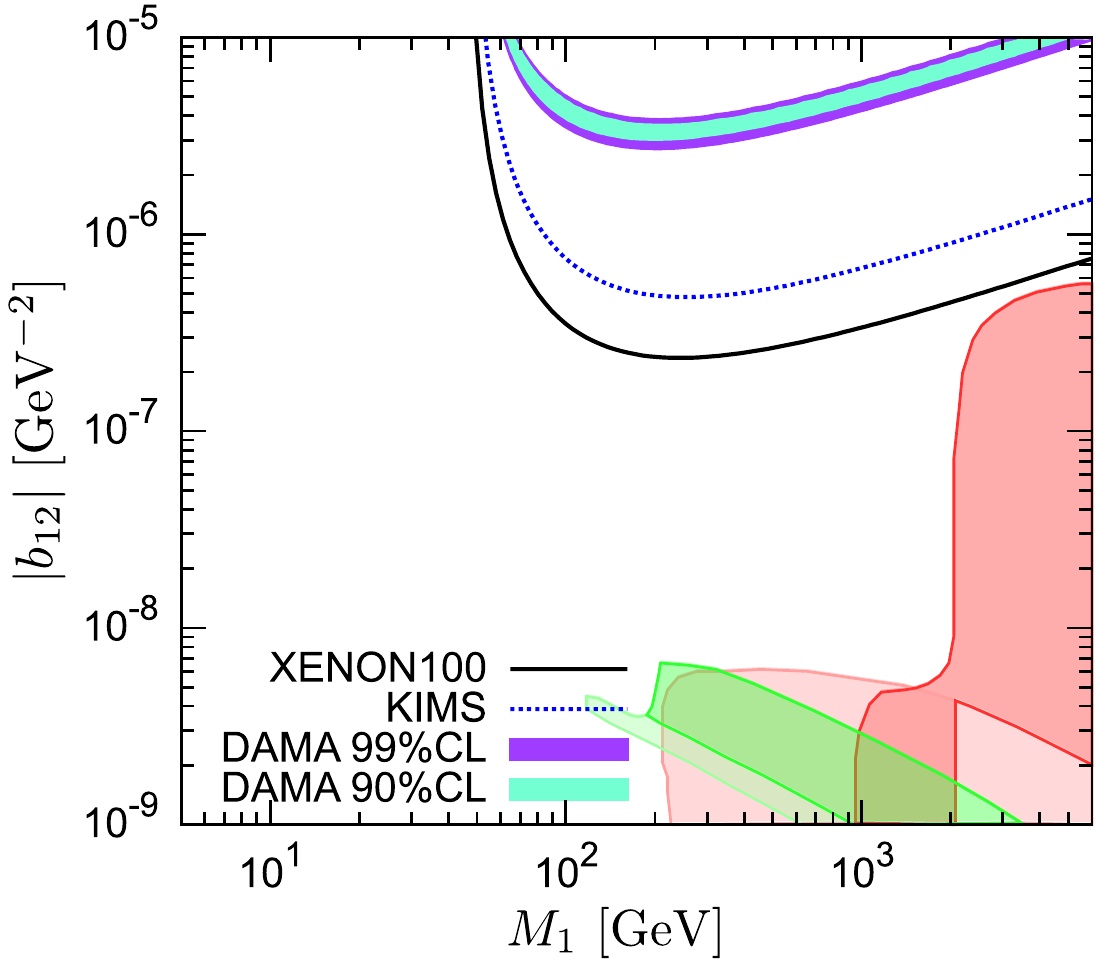} 

\mycaption{Bounds from XENON100, KIMS and allowed regions for DAMA in
  the $(M_1,|b_{12}|)$ plane (charge-charge interaction).  The mass
  difference $\delta$ is taken as $0~\mathrm{keV}$ (the left top
  panel), $40~\mathrm{keV}$ (the right top panel), $80~\mathrm{keV}$
  (the left bottom panel) and $120~\mathrm{keV}$ (the right bottom
  panel). The shaded regions correspond to the values of $b_{12}$
  predicted in the allowed parameter space of the model, as shown in
  Fig.~\ref{fig:relic}, with the same color shading for different
  values of the ratio $M_\eta/M_1$.}
\label{fig:xenon-kims-dama}
\end{center}
\end{figure}

\begin{figure}[t!]
\begin{center}
\includegraphics[scale=0.65]{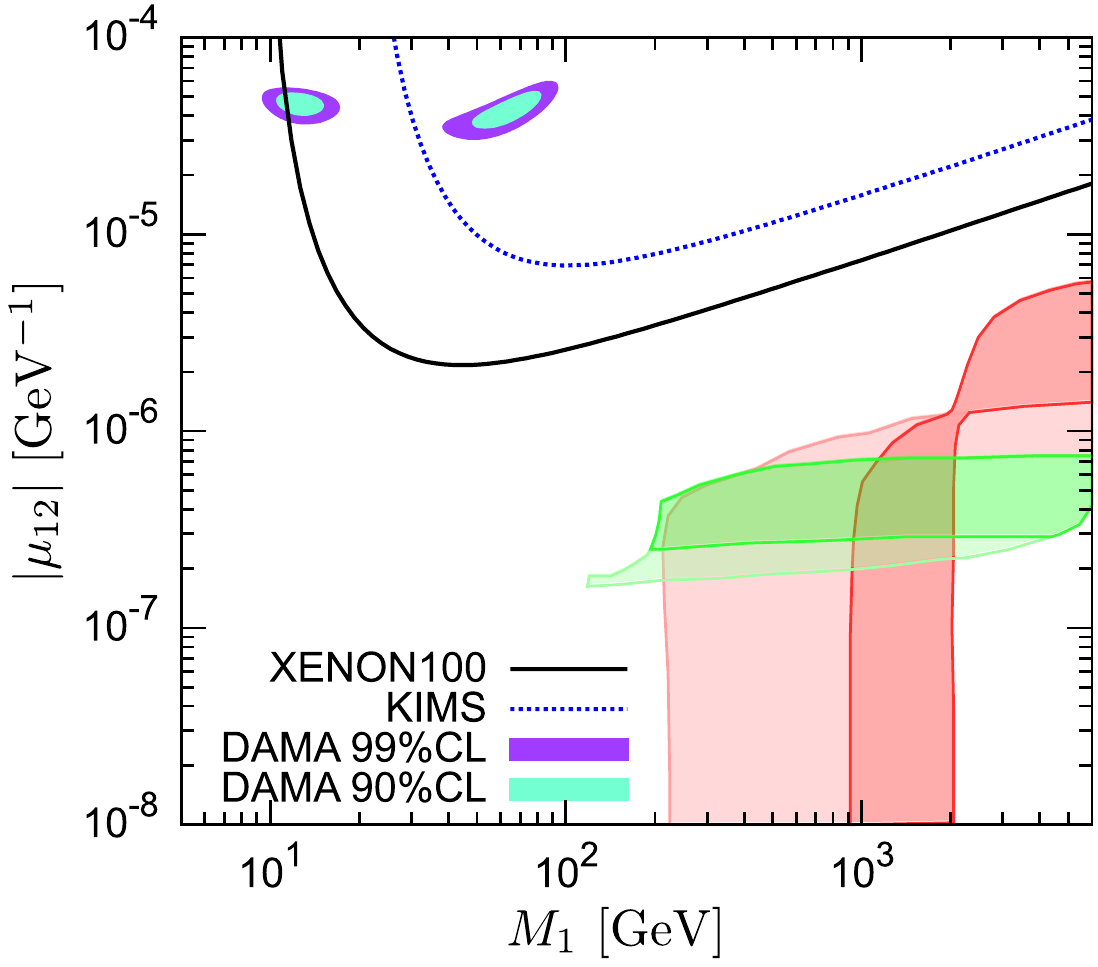}\qquad
\includegraphics[scale=0.65]{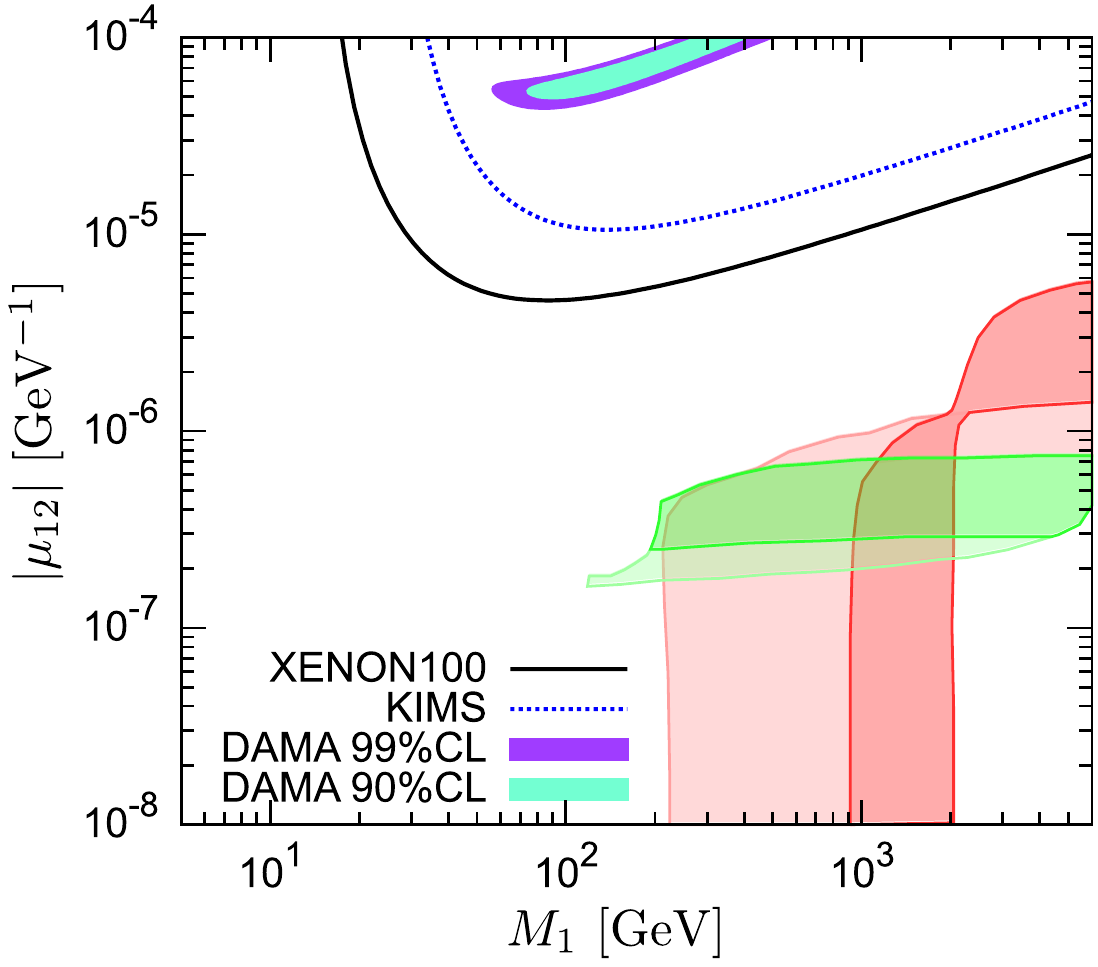}\\\quad\\
\includegraphics[scale=0.65]{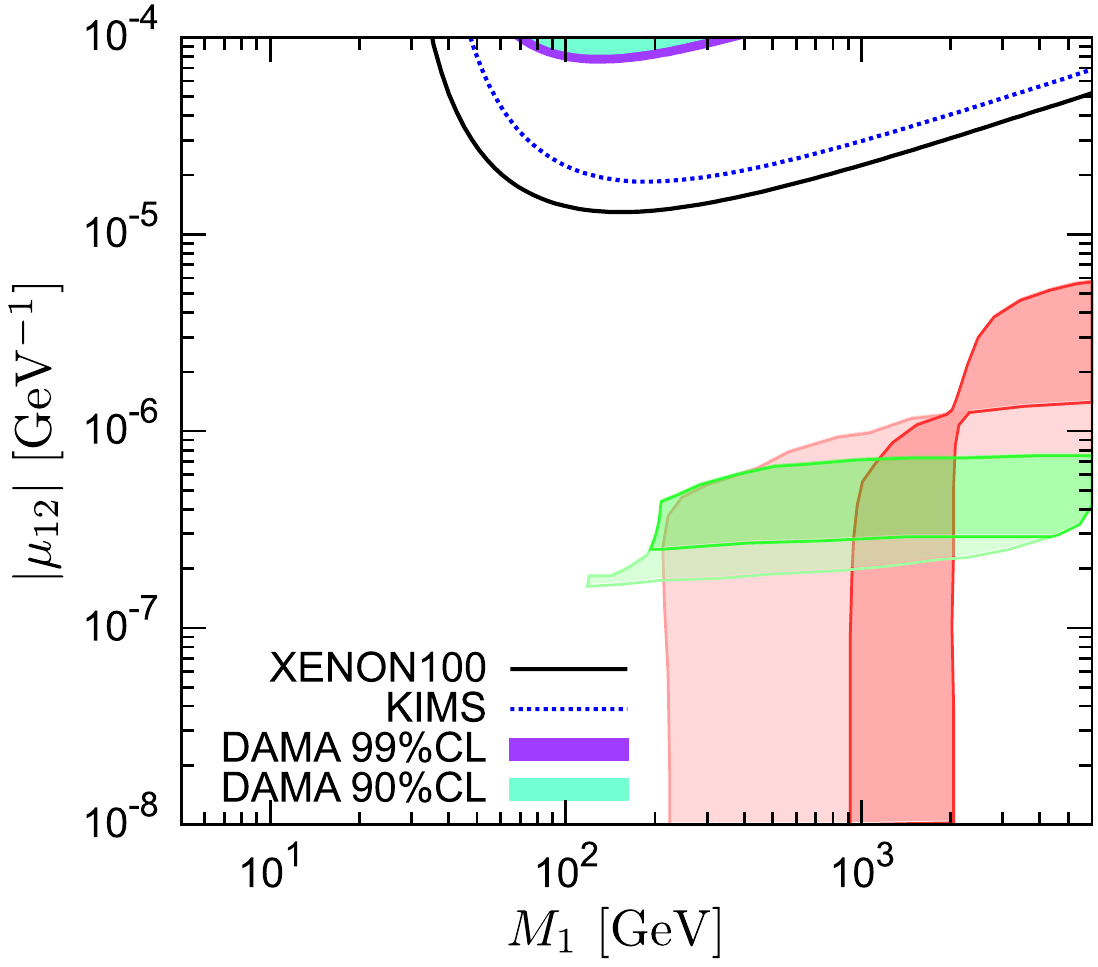}\qquad
\includegraphics[scale=0.65]{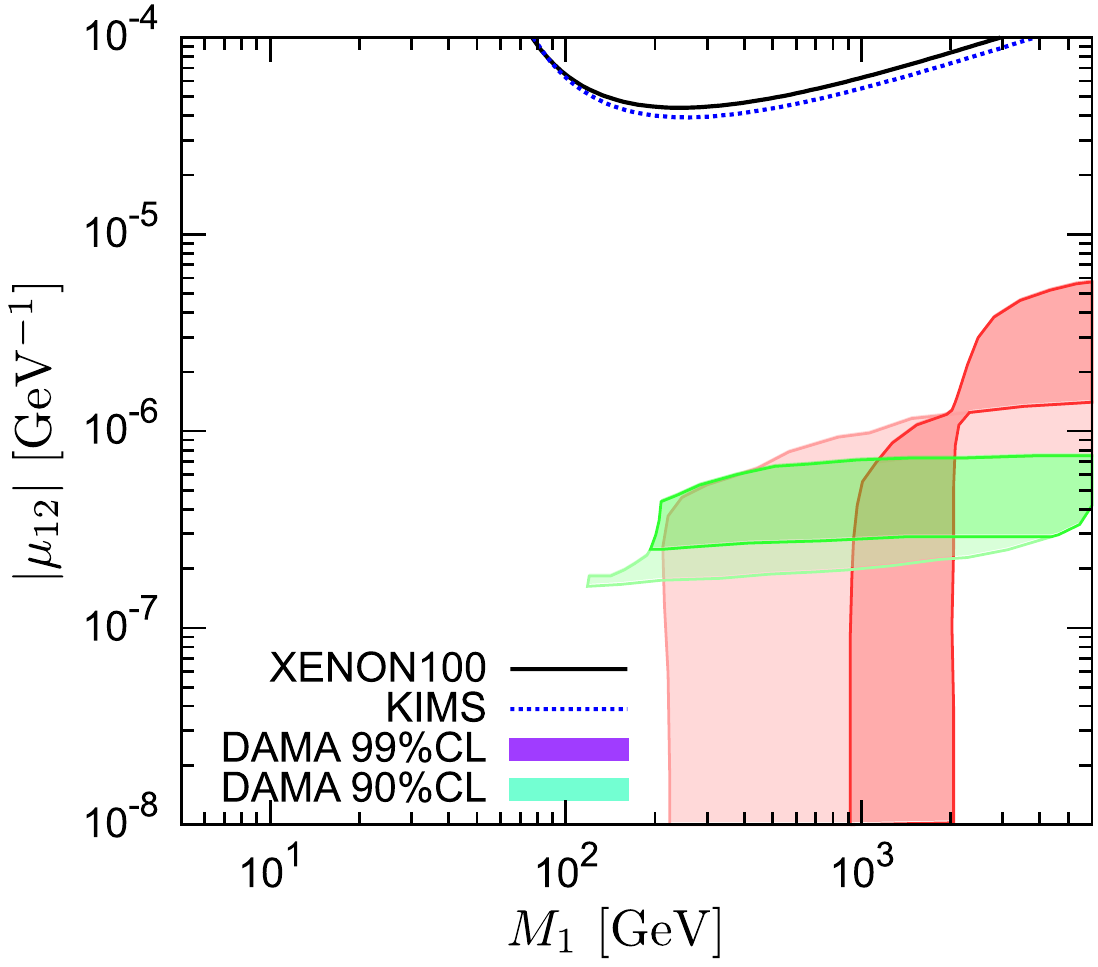}

\mycaption{Bounds from XENON100, KIMS and allowed regions for DAMA in
  the $(M_1,|\mu_{12}|)$ plane (dipole-charge and dipole-dipole
  interaction). The mass difference $\delta$ is taken as
  $0~\mathrm{keV}$ (the left top panel), $40~\mathrm{keV}$ (the right
  top panel), $80~\mathrm{keV}$ (the left bottom panel) and
  $120~\mathrm{keV}$ (the right bottom panel). The shaded regions
  correspond to the values of $\mu_{12}$ predicted in the allowed
  parameter space of the model, as shown in Fig.~\ref{fig:relic}, with
  the same color shading for different values of the ratio
  $M_\eta/M_1$.}
\label{fig:xenon-kims-dama2}
\end{center}
\end{figure}

In order to derive constraints on the model we calculate the total
event rate for XENON100 and KIMS in the energy range given in
Tab.~\ref{tab:qf} and require that the predicted rate is less than
$0.0017$, $0.0098~\mathrm{kg^{-1}day^{-1}}$ for XENON100 and KIMS,
respectively. The upper bounds are obtained from the observed
$3$ events with $3\sigma$ of the statistical error in
the $48~\mathrm{kg}$ fiducial volume during $100.9$ live days exposure
in the signal region for XENON100~\cite{Aprile:2011hi}, and from
ref.~\cite{kims-taup} for KIMS. 
For DAMA we perform a $\chi^2$ fit to the
modulation amplitude in bins of observed scintillation energy between
2 and 8~keVee. In Fig.~\ref{fig:xenon-kims-dama} and
Fig.~\ref{fig:xenon-kims-dama2} we show the bounds from XENON100, KIMS
and allowed regions from DAMA for the coefficients $b_{12}$ (see
Eq.~(\ref{eq:zz})) and $\mu_{12}$ (see Eqs.~(\ref{eq:dz}),
(\ref{eq:dd})), respectively.  These bounds are compared to the regions
as predicted in the model according to Eqs.~(\ref{eq:a12}),
(\ref{eq:c12}), (\ref{eq:b12}) for $b_{12}$ and Eq.~(\ref{eq:mu12}) for
$\mu_{12}$. The colored regions correspond to the regions shown in
Fig.~\ref{fig:relic}, satisfying constraints from neutrino masses and
mixing, charged lepton-flavour violation, the relic DM density, and
perturbativity.  The ratio $M_\eta/M_1$ is taken in the range $1 \le
M_\eta/M_1 \leq {9.8}$, with the same color shading as in
Fig.~\ref{fig:relic}.  There is no allowed parameter space for
$M_\eta/M_1 \gtrsim 9.8$, as discussed earlier.

We observe that the values of $|b_{12}|$ and $|\mu_{12}|$ obtained in this
model are too small to account for the signal in DAMA. For very small mass
splittings $\delta$ between $N_1$ and $N_2$ some regions of the parameter
space are excluded by XENON100 data. The constraints become weaker for
larger $\delta$, since increasing inelasticity suppresses the scattering
event rate. 
Relatively large values of $|b_{12}|$ are obtained for close to
degenerate $N_1$ and $\eta$, $M_\eta/M_1\lesssim 1.05$ (dark-red
region), because of the behavior of the function $I_{\mathrm{a}}(x,y)$
near $x=1$, where $I_{\mathrm{a}}(x,y)\sim y^{-1}$ and $y =
m_\alpha^2/M_\eta^2$ is small. The region excluded by XENON100 for $M_1
\simeq M_\eta \sim 2$~TeV becomes allowed for $\delta \gtrsim
120$~keV (bottom-right panel). By comparing Figs.~\ref{fig:xenon-kims-dama} and
\ref{fig:xenon-kims-dama2} we observe that the model predicts values
of $|\mu_{12}|$ too small to be tested by current direct detection data.
The enhancement for the transition magnetic moment
$|\mu_{12}|$ for $M_\eta/M_1\lesssim 1.05$ is less than for $|b_{12}|$
due to a different behavior of the loop functions
$I_{\mathrm{a}}(x,y)$ and $I_{\mathrm{m}}(x,y)$. 

\begin{figure}[t!]
\begin{center}
\includegraphics[scale=0.65]{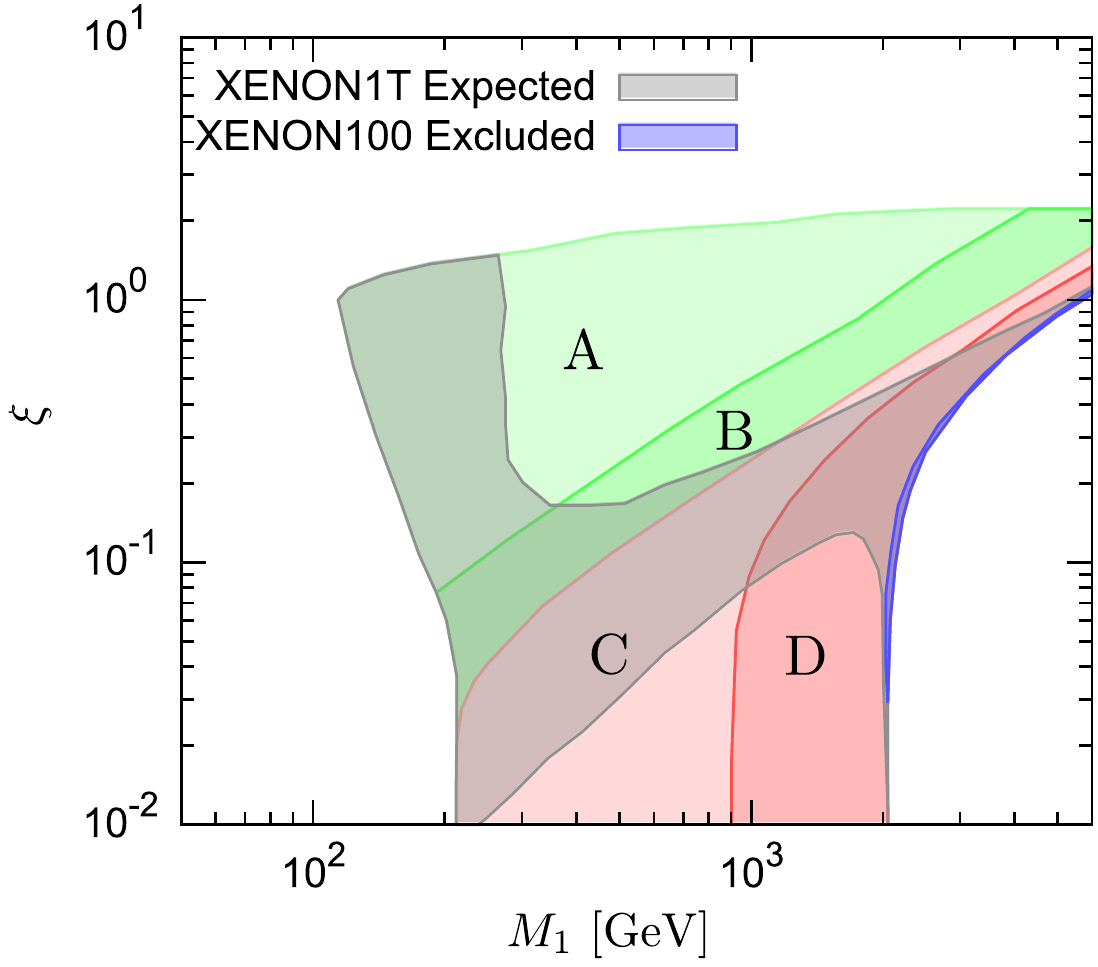}\qquad
\includegraphics[scale=0.65]{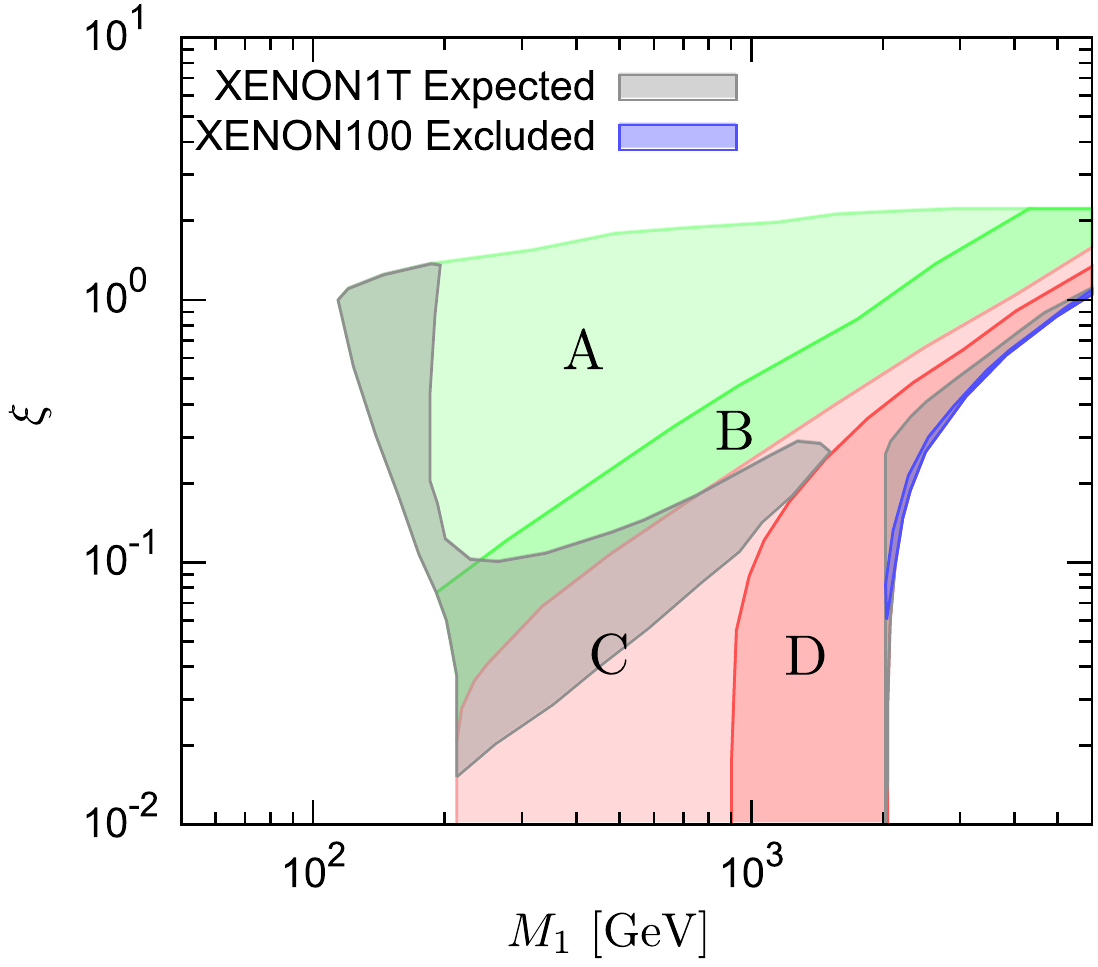}\\\quad\\
\includegraphics[scale=0.65]{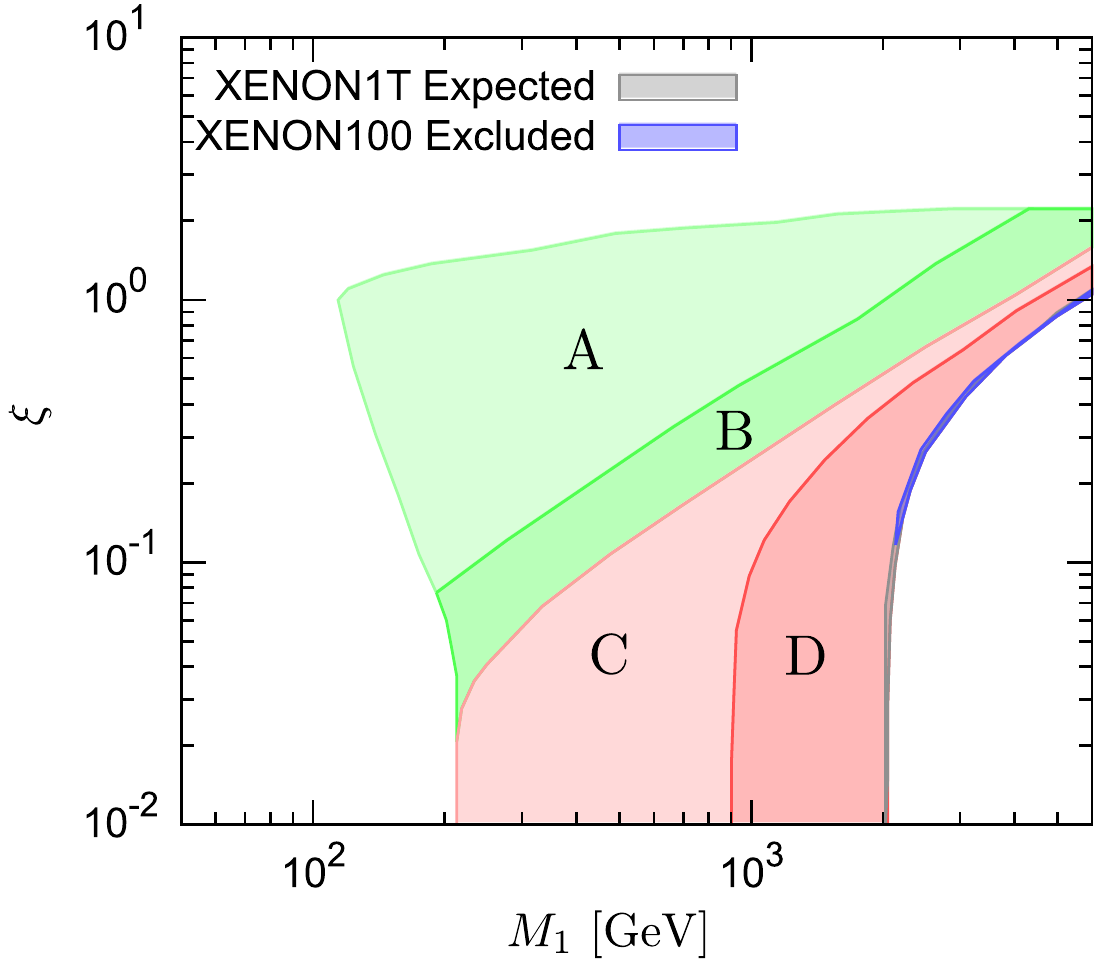}\qquad
\includegraphics[scale=0.65]{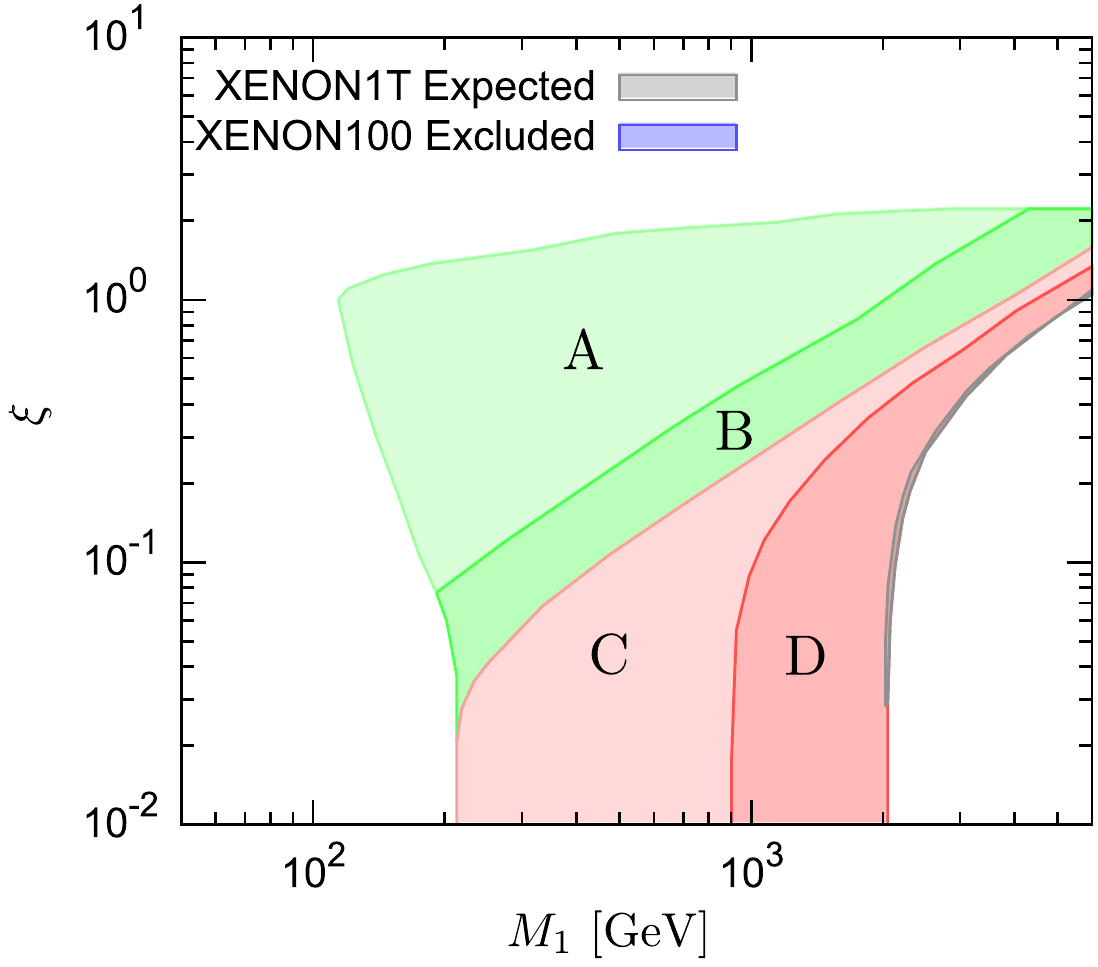}
\mycaption{Same as
Fig.~\ref{fig:relic} with constraints from XENON100 (blue) and sensitivity from
XENON1T (gray) overlayed. We assume $\delta=$0 (left top), 
40~keV (right top), 80~keV (left bottom),
 120~keV (right bottom).} 
\label{fig:ykw}
\end{center}
\end{figure}

We conclude that current data from XENON100 start to exclude some
parameter space of the model, in case of degenerate configurations
$M_1 \simeq M_2 \simeq M_\eta \sim$~few TeV. In
  Fig.~\ref{fig:ykw} we show the regions excluded from XENON100
  overlayed to the globally allowed regions from Fig.~\ref{fig:relic}
  as dark blue, by translating the the $|b_{12}|$ constraint into a
  bound on $\xi$. Furthermore we show in Fig.~\ref{fig:ykw} the
  estimated sensitivity for XENON1T. Using the sensitivity for the
  elastic WIMP-nucleon scattering cross section from
  ref.~\cite{xenon1t} we estimate that XENON1T will constrain the
  event rate to be less than
  $1.59\times10^{-5}~[\mathrm{kg^{-1}day^{-1}}]$.  (We assume the same
  nuclear recoil energy range as for XENON100.) Then we compare this
  number to the event rate induced in the model assuming several
  values for the mass splitting $\delta$. From Fig.~\ref{fig:ykw} we
  find that for $\delta \lesssim 40$~keV future data from the XENON1T
  experiment~\cite{xenon1t} will dig deeply into the allowed parameter
  region of the model. For 40~keV~$\lesssim \delta \lesssim 120$~keV
  the degenerate region $M_1 \simeq M_2 \simeq M_\eta \sim$~few TeV
  will be tested. We note however, that no signal is guaranteed for
direct detection. In the $N_1 - \eta$ co-annihilation region (dark-
and light-red regions, where $M_\eta / M_1 < 1.2$) no lower bound on
the parameter $|\xi|$ is obtained, leading to arbitrarily small values
of $|b_{12}|$ and $|\mu_{12}|$, which implies a vanishing signal in
direct detection experiments.


\subsection{Monochromatic Photon from the Decay of $N_2$}

\begin{figure}[t]
\begin{center}
\includegraphics[scale=1]{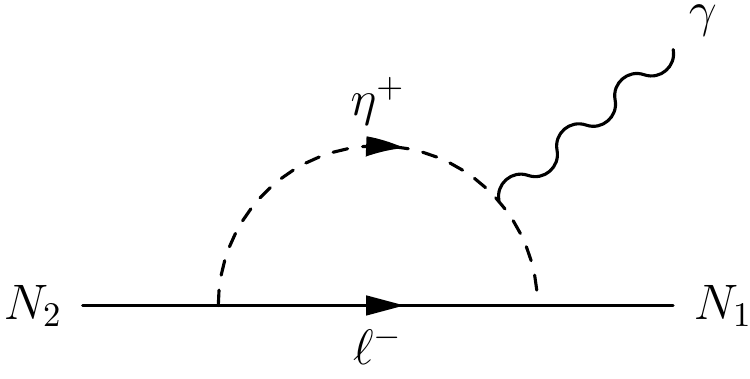}
\quad
\includegraphics[scale=1]{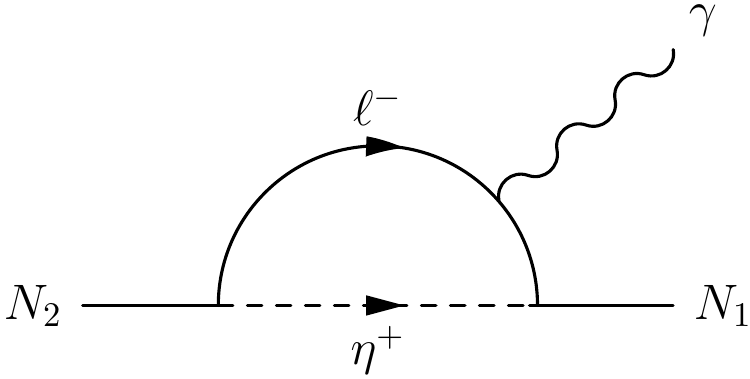}
\mycaption{Decay process of $N_2$.}
\label{fig:decay}
\end{center}
\end{figure}

The excited DM state $N_2$ decays to $N_1$ and a photon
through the transition magnetic moment. 
The diagrams of the decay process are shown in Fig.~\ref{fig:decay} and
the decay width is calculated as 
\begin{equation}
\Gamma(N_2\to N_1\gamma)=
\frac{\mu_{12}^2}{\pi}\delta^3 \,.
\end{equation}
Notice that the effective interaction $b_{12}$ does not contribute to
the decay width since the emitted photon is on-shell.  The decay of
$N_2$ produces a monochromatic photon of energy $E_\gamma\simeq
\delta$.  If the decay happens inside a DM detector this monochromatic
photon would contribute to the electromagnetic event rate. Although
typically such events are rejected in order to search for nuclear
recoils one may be able to place constraints on the model by requiring
that the electromagnetic event rate induced by the decay of $N_2$ has
to be less than the observed rate. A similar mechanism has been used in 
ref.~\cite{Feldstein:2010su} in order to explain the DAMA modulation signal.

Following ref.~\cite{Feldstein:2010su} we estimate the photon induced
event rate in the model under consideration for the XENON100
experiment.  The excited state $N_2$ is produced by the inelastic
scattering with nuclei inside the Earth which is composed of various
elements such as Fe, O and Si.  The event rate in XENON100 is given by
\begin{equation}
\frac{dR_{\gamma}}{dE_R}=\frac{\rho_{\odot}}{M_1\rho_{\mathrm{Xe}}}
\sum_{i = \mathrm{nuclei}}\int_{v>v_{\mathrm{min}}}{d^3v}
\frac{d\sigma_i}{dE_R}vf(\bm{v})\int_{\mathrm{Earth}}d^3r \,
n_i(\bm{r})P(\bm{r},v),
\label{eq:gamma}
\end{equation}
where $\rho_{\mathrm{Xe}}$ is the mass density of the XENON detector
$21.9~\mathrm{g/cm^3}$, $\sigma_i$ is the total inelastic scattering
cross section which includes the charge-charge, dipole-charge and
dipole-dipole interactions, and $n_i(\bm{r})$ is the number density
for the given nucleus $i$ inside the Earth. Note that $E_R$ is the
nuclear recoil energy in the $N_1 + A \to N_2 + A$ scattering process.
The contribution of the dipole-dipole interaction is much smaller than
the ones from the charge-charge and dipole-charge interactions since
the fraction of isotopes with a sizable magnetic moment in the Earth
is less than a few \%.  In Eq.~(\ref{eq:gamma}), $P(\bm{r}, v)$ is the
probability that an $N_2$ which is produced by the scattering of DM
with velocity $v$ at the position $\bm{r}$ decays inside the
XENON100 detector. It is given by
\begin{equation}
P(\bm{r},v)=\frac{1}{4\pi(\bm{r}-\bm{r}_{\mathrm{Xe}})^2}\frac{\Gamma}{v_f}
e^{-\Gamma|\bm{r}-\bm{r}_{\mathrm{Xe}}|/v_f},
\end{equation}
where $v_f=\sqrt{v^2-2(\delta+E_R)/M_1}$ is the velocity of the
produced $N_2$, and $\bm{r}_{\mathrm{Xe}}$ is the position of XENON
detector on the Earth.  The total gamma event rate $R_{\gamma}$ is
obtained by integrating Eq.~(\ref{eq:gamma}) over the recoil energy
$E_R$. 

In order to obtain a rough estimate of the induced event rate we 
introduce some approximations. We
use the averaged number density of the elements in the Earth
$\overline{n}\simeq 9.85\times10^{22}~\mathrm{cm^{-3}}$, the averaged
atomic number $\overline{Z}\simeq29.9$ and the averaged magnetic
moment of nuclei $\overline{\mu}_A/\mu_N\simeq3.46\times10^{-2}$, which
are calculated by taking into account the structure of the Earth such
as the crust, mantle and core~\cite{Feldstein:2010su}.  Replacing 
$n_i(\bm{r})$ by its average, it can be pulled out of the $r$-integral in 
Eq.~(\ref{eq:gamma}) and the
integration is performed analytically:
\begin{equation}
\int_{\mathrm{Earth}}d^3rP(\bm{r},v)=
\frac{1}{2}\left[\frac{v_f}{2\Gamma r_{\oplus}}
\left(e^{-2\Gamma{r_{\oplus}}/v_f}-1\right)+1\right],
\end{equation}
where $r_{\oplus}=6.4\times10^{6}~\mathrm{m}$ is the radius of the Earth.
The remaining integrations over $v$ and $E_R$ are done numerically.

With this approximation we estimate the total predicted event rate in
XENON100 for typical parameters of the model. We find that the maximal
rate is approximately $R_{\gamma}^{\mathrm{max}}\simeq
2.0\times10^{-7}~\mathrm{kg}^{-1}\mathrm{day}^{-1}$, when
$\delta\simeq 40~\mathrm{keV}$ and $M_1/M_\eta=1$. This result should
be compared with 22 events obtained in the electromagnetic band in the
$40~\mathrm{kg}$ fiducial volume during $11.17$ live days exposure in the
DM search window by XENON100~\cite{Aprile:2010um}. Hence, since the
predicted event rate is several orders of magnitude smaller we
conclude that the monochromatic photon from the $N_2$ decay will not
lead to any observable signal in DM direct detection experiments.


\section{Summary and Conclusions}

We have considered a model proposed by Ma~\cite{Ma:2006km}, providing
an economical extension of the Standard Model to accommodate neutrino
masses and DM. The Standard Model is extended by three fermion
singlets $N_i$ (``right handed neutrinos'') and an inert scalar
doublet $\eta$, where the new particles transform odd under a
$\mathbb{Z}_2$ symmetry, making the lightest of them a stable DM
candidate. In our case $N_1$ is the DM particle. We investigate the
parameter space of the model consistent with neutrino masses and
mixings, bounds on charged lepton-flavour violation, perturbativity,
and the correct relic DM abundance due to the thermal freeze-out
mechanism. We find that in order to obtain the correct relic DM
abundance co-annihilations are always important, either between the
two lightest fermion singlets $N_1$ and $N_2$, or between $N_1$ and
the inert doublet $\eta$. 

In this model DM has no direct couplings to quarks and gluons. Despite
this leptophilic nature of DM, scattering off nuclei is possible at
1-loop level by photon exchange. We have calculated the relevant loop
processes in an effective field theory approach. One obtains effective
charge-charge, dipole-charge, and dipole-dipole interactions between
DM and nuclei, leading to a non-vanishing scattering rate in DM direct
detection experiments. The scattering is inelastic and in order to
obtain a sizable scattering rate $N_1$ and $N_2$ have to be highly
degenerate, with mass differences $\delta$ less than few 100~keV. This
is consistent with the need for co-annihilations to obtain the correct
relic abundance. Although the scattering cross section in this model
is too small to account for the DAMA annual modulation signal, we find
that for mass differences $\delta \lesssim 120$~keV current data from
the XENON100 experiment start to exclude certain regions of the
parameter space. The predicted event rate for XENON100 is dominated by
the charge-charge interaction. Future data, for example from XENON1T,
will significantly dig into the allowed parameter space and provide a
stringent test for the model provided $\delta$ is small enough.

\subsection*{Note added.}

After this work has been completed the Daya Bay reactor experiment
released their data \cite{An:2012eh}, establishing a non-zero value of
$\theta_{13}$ at more than $5\sigma$ with $\sin^22\theta_{13} =
0.092\pm 0.016\pm 0.005$. For non-zero values of $\theta_{13}$
additional contributions to $\mu\to e \gamma$ are induced, providing
further constraints on the model. Daya Bay data imply $\sin\theta_{13}
> 0.1$ at $3\sigma$, which constrains DM masses around the TeV scale,
see Fig.~\ref{fig:relic}.


\section*{Acknowledgments}
T.T.\ is supported by Young Researcher Overseas Visits
Program for Vitalizing Brain Circulation Japanese in JSPS. 
D.S.\ is supported by the International Max Planck Research School for
Precision Tests of Fundamental Symmetries. 
T.T.\ would like to thank Daijiro Suematsu for useful
comments and the Particle and Astroparticle Physics group at MPIK Heidelberg. 
The numerical calculations were partially carried out on SR16000 at YITP
in Kyoto University.


\appendix
\section*{Appendix A}
\subsection*{Explicit Functions for the Effective Interactions}
Here we give the explicit functions for the effective interactions. 
The functions $I_{\mathrm{a}}(x,y)$ and
$I_{\mathrm{m}}(x)$ are given as follows, 
\begin{eqnarray}
I_{\mathrm{a}}(x,y)\!\!\!&=&\!\!\!\frac{1}{3}\int_0^1\frac{3u^2-6u+1}{xu^2-(1+x-y)u+1},\\
I_{\mathrm{m}}(x,y)\!\!\!&=&\!\!\!-\int_0^1\frac{u(1-u)}{xu^2-(1+x-y)u+1}.
\end{eqnarray}
The analytic formulas of these integrations are given as follows. 
\begin{itemize}
\item[(i)]
If $(1+x-y)^2-4x>0$, 
\end{itemize}
\begin{eqnarray}
I_\mathrm{a}\left(x,y\right)\!\!\!&=&\!\!\!
\frac{1}{x}\left[1+\frac{3A_{+}^2-6A_{+}+1}{3(A_{+}-A_{-})}\log\left|\frac{A_{+}-1}{A_{+}}\right|
-\frac{3A_{-}^2-6A_{-}+1}{3(A_{+}-A_{-})}\log\left|\frac{A_{-}-1}{A_{-}}\right|\right],\\ 
I_{\mathrm{m}}(x,y)\!\!\!&=&\!\!\!
\frac{1}{x}\left[1+\frac{A_+(A_{+}-1)}{A_{+}-A_{-}}
\log\left|\frac{A_{+}-1}{A_{+}}\right|-\frac{A_{-}(A_{-}-1)}{A_{+}-A_{-}}
\log\left|\frac{A_{-}-1}{A_{-}}\right|\right].
\end{eqnarray}
\begin{itemize}
\item[(ii)]
If $(1+x-y)^2-4x=0$,
\end{itemize}
\begin{eqnarray}
I_\mathrm{a}\left(x,y\right)\!\!\!&=&\!\!\!
\frac{1}{x}\left[1+2(A_0-1)\log\left|\frac{A_{0}-1}{A_{0}}\right|
+\frac{3A_0^2-6A_0+1}{3A_0(A_0-1)}\right],\\ 
I_{\mathrm{m}}(x,y)\!\!\!&=&\!\!\!
\frac{1}{x}\left[2+(2A_{0}-1)\log\left|\frac{A_{0}-1}{A_{0}}\right|\right]. 
\end{eqnarray}
\begin{itemize}
\item[(iii)]
If $(1+x-y)^2-4x<0$,
\end{itemize}
\begin{eqnarray}
I_{\mathrm{a}}(x,y)\!\!\!&=&\!\!\!
\frac{1}{x}\left[1+\frac{B_{+}+B_{-}-2}{2}\log\left|\frac{(B_{+}-1)^2+(B_{-}-1)^2}{B_{+}^2+B_{-}^2}
\right|\right.\nonumber\\
&&\qquad\left.+\frac{6(B_{+}-1)(B_{-}-1)-4}{3(B_{+}-B_{-})}
\mathrm{Tan}^{-1}\left(\frac{B_{+}-B_{-}}{B_{+}^2+B_{-}^2-B_{+}-B_{-}}\right)\right],\\
I_{\mathrm{m}}(x,y)\!\!\!&=&\!\!\!
\frac{1}{x}\left[1+\frac{B_{+}+B_{-}-1}{2}\log\left|\frac{(B_{+}-1)^2+(B_{-}-1)^2}{B_{+}^2+B_{-}^2}
\right|\right.\nonumber\\
&&\qquad\left.+\frac{(2B_{+}-1)(2B_{-}-1)-1}{2(B_{+}-B_{-})}
\mathrm{Tan}^{-1}\left(\frac{B_{+}-B_{-}}{B_{+}^2+B_{-}^2-B_{+}-B_{-}}\right)\right].
\end{eqnarray}
$A_{\pm}$, $A_{0}$ and $B_{\pm}$ are defined as 
\begin{eqnarray}
A_{\pm}\!\!\!&\equiv&\!\!\!\frac{1+x-y\pm\sqrt{(1+x-y)^2-4x}}{2x},\nonumber\\
A_{0}\!\!&\equiv&\!\!\!\frac{1+x-y}{2x},\nonumber\\
B_{\pm}\!\!\!&\equiv&\!\!\!\frac{1+x-y\pm\sqrt{4x-(1+x-y)^2}}{2x}.\nonumber
\end{eqnarray}
The function $I_{\mathrm{c}}(x,y)$ is the same as
$I_{\mathrm{m}}(x,y)$. 
These functions are continuous and smooth for $0\leq x,y\leq1$.
For $0\simeq y\ll x\ll 1$, these functions approach to 
\begin{eqnarray}
I_{\mathrm{a}}(x,y)\!\!\!&\to&\!\!\!\:\:\:\:\!
\frac{1}{2}+\frac{2}{3}\log{y},\\
I_{\mathrm{m}}(x,y)\!\!\!&\to&\!\!\!
-\frac{1}{2}.
\end{eqnarray}
Therefore, the obtained parameters $|b_{12}|$ and $|\mu_{12}|$ at lowest
order agree with the result of ref.~\cite{Agrawal:2011ze} where
the parameter $\lambda^2$ in ref.~\cite{Agrawal:2011ze} corresponds to
$\mathrm{Im}(h_{\alpha 2}^*h_{\alpha 1})/2$ in our notation. 
The difference of the relative sign comes from the denifition of the
effective operators.

\end{document}